\documentclass[preprint,aps,jcp,onecolumn,superscriptaddress]{revtex4-1}

\usepackage[paperheight=11in,paperwidth=8.25in,margin=0.67in]{geometry}

\usepackage[T1]{fontenc} 
\usepackage[version=3]{mhchem} 
\usepackage{mathpazo}

\usepackage{amsfonts}
\usepackage{amsmath}
\usepackage{booktabs}
\usepackage{tikz}
\usepackage{threeparttable}
\usepackage{multirow}
\usepackage{appendix}
\usepackage{amssymb}
\usepackage{graphicx}
\usepackage{subfigure}
\usepackage{enumerate}
\usepackage{colortbl}
\usepackage{framed}
\usepackage{verbatim}
\usepackage{amsbsy}
\usepackage{bm}

\usepackage[normalem]{ulem}
\usepackage{float}
\usepackage[linesnumbered,boxed]{algorithm2e}
\usepackage{algorithm2e}

\usepackage{simplewick} 
\usepackage{bbm}

\newcommand{\sgn}{ {\rm{sgn}} }

\newfloat{Algorithm}{htbp}{alg}
\floatname{Algorithm}{Algorithm}

\newcounter{exe}[figure]
\newcommand{\iexe}{\refstepcounter{exe}\the\value{exe}:}

\begin{document}

\title{Perspective: Essentials of Relativistic Quantum Chemistry}

\author{Wenjian Liu}\email{liuwj@sdu.edu.cn}
\affiliation{Qingdao Institute for Theoretical and Computational Sciences, Shandong University, Qingdao, Shandong 266237, P. R. China}

\begin{abstract}
Relativistic quantum chemistry has evolved into a fertile and large field and is now becoming an integrated part of mainstream chemistry.
Yet, given the much-involved physics and mathematics (as compared with nonrelativistic quantum chemistry), it is still necessary to
clean up the essentials underlying the relativistic electronic structure theories and methodologies (such that uninitiated readers can
pick up quickly the right ideas and tools for further development or application)
and meanwhile pinpoint future directions of the field. To this end,
the three aspects of electronic structure calculations, i.e., relativity, correlation, and QED,
will be highlighted.
\end{abstract}

\maketitle

\section{Introduction: ABC of Relativistic Quantum Mechanics}
As one of the two pillars of modern physics,
the theory of special relativity was founded by Einstein in 1905 \cite{Einstein1905}. Among others, the best known energy-mass relation
\begin{equation}
E=\gamma m c^2\label{MassE}
\end{equation}
is most relevant for our purpose. Here, $m$ is the rest mass of a particle moving with velocity $\boldsymbol{v}$, while $\gamma=(1-\frac{\boldsymbol{v}^2}{c^2})^{-1/2}$ is the Lorentz factor, with $c$ being the constant velocity of light.
This relation can be converted to
\begin{equation}
E^2=c^2\boldsymbol{p}^2+m^2c^4\label{Esquare}
\end{equation}
via the very definition of the momentum $\boldsymbol{p}=\gamma m \boldsymbol{v}$. In fact, the energy-momentum relation \eqref{Esquare} is more fundamental than the energy-mass relation \eqref{MassE},
since the former applies to both massive and massless particles whereas the latter applies only to massive particles. Moreover,
relation (1) is merely the positive-energy part of the square root of the right-hand side of relation \eqref{Esquare}, i.e., $E=\pm\sqrt{c^2\boldsymbol{p}^2+m^2c^4}$.
In view of the correspondence principle, i.e.,
\begin{equation}
E\rightarrow i\hbar\frac{\partial}{\partial t},\quad \boldsymbol{p}\rightarrow -i\hbar\boldsymbol{\nabla},\label{Correspondence}
\end{equation}
relation \eqref{Esquare} can be mapped, as done by Klein and Gordon in 1926\cite{Klein:26,Gordon:26}, to a first-quantized wave equation,
\begin{equation}
(\frac{1}{c^2}\frac{\partial^2}{\partial t^2}-\boldsymbol{\nabla}^2+k^2)\psi_{\mathrm{KG}}(x)=0,\quad k=\frac{mc}{\hbar},\quad x=\boldsymbol{r}t,\label{KGeq}
\end{equation}
which treats space and time on the same footing and is manifestly Lorentz covariant.
However, it is second order in time, which is fundamentally different from
the Schr\"odinger equation that is first order in time. As a result, the norm-conserving density\cite{greiner1990relativistic}, $\rho_{\mathrm{KG}}=\frac{i\hbar}{2mc^2}(\psi_{\mathrm{KG}}^*\frac{\partial\psi_{\mathrm{KG}}}{\partial t}-\psi_{\mathrm{KG}}\frac{\partial\psi_{\mathrm{KG}}^*}{\partial t})$,
is not positive definite (since $\psi_{\mathrm{KG}}$ and $\frac{\partial\psi_{\mathrm{KG}}}{\partial t}$ are independent of each other and can have arbitrary values at a given time $t$) and hence cannot be interpreted as a probability density.
Because of this, the Klein-Gordon equation \eqref{KGeq} was not regarded to be physically meaningful until Pauli and Weisskopf\cite{Pauli1934} recognized, through a theoretical exercise\cite{HistoryQED}, that
it is a relativistic wave equation for spin-0, charged and massive particles, which were
discovered to be $\pi^+$ and $\pi^-$ mesons in the late 1940s (NB: $\rho_{\mathrm{KG}}$
multiplied by charge $q$ can be reinterpreted as a charge density which can be either positive or negative).
A relativistic first-quantized wave equation that is first order in time \emph{and} in space was first proposed by Dirac in 1928\cite{Dirac1928a,Dirac1928b},
by noticing that the energy-momentum relation \eqref{Esquare} can be written as
\begin{eqnarray}
E^2&=&D_0^2,\quad D_0=c\boldsymbol{\alpha}\cdot\boldsymbol{p}+\beta mc^2=D_0^\dag\label{Dop}\\
&=&c^2\sum_{i=x,y,z} \alpha_i^2p_i^2+\beta^2m^2c^4+\sum_{i=x,y,z}[\beta, \alpha_i]_+ p_imc^3+\frac{c^2}{2}\sum_{i\ne j}^3 [\alpha_i,\alpha_j]_+p_ip_j\\
&=&c^2\boldsymbol{p}^2+m^2c^4,
\end{eqnarray}
provided that the following conditions hold
\begin{eqnarray}
\beta^\dag=\beta,\quad \alpha_i^\dag=\alpha_i,\quad \beta^2=\alpha_i^2=1,\quad [\beta, \alpha_i]_+= [\alpha_i,\alpha_j]_+=0 \mbox{ }(i\ne j).
\end{eqnarray}
It is then not a difficult math to figure out the explicit, lowest-dimensional matrix expressions for $\beta$ and $\boldsymbol{\alpha}$:
\begin{eqnarray}
\beta=\begin{pmatrix}1&0\\ 0&-1\end{pmatrix},\quad \boldsymbol{\alpha}=\begin{pmatrix}0&\boldsymbol{\sigma}\\ \boldsymbol{\sigma}&0\end{pmatrix},\label{DiracMat}
\end{eqnarray}
where $\boldsymbol{\sigma}$ is the vector of the $2\times 2$ Pauli spin matrices,
\begin{eqnarray}
\sigma_x=\begin{pmatrix}0&1\\ 1&0\end{pmatrix},\quad \sigma_y=\begin{pmatrix}0&-i\\ i&0\end{pmatrix},\quad \sigma_z=\begin{pmatrix}1&0\\ 0&-1\end{pmatrix}.\label{PauliMat}
\end{eqnarray}
Recognizing the function $D_0$ \eqref{Dop} as the Hamiltonian,
the correspondence principle \eqref{Correspondence} then leads immediately to the free-particle Dirac equation
\begin{equation}
i\hbar\frac{\partial}{\partial t}\psi(x)=D_0\psi(x).\label{fp-DEQ}
\end{equation}
At variance with the scalar form of $\psi_{\mathrm{KG}}$, the wave function $\psi$ of the Dirac equation \eqref{fp-DEQ}
is a bispinor with four components, i.e.,
\begin{equation}
\psi(x)=\begin{pmatrix}\psi^L(x) \\ \psi^S(x)\end{pmatrix}=\begin{pmatrix}\psi^{L\alpha}(x) \\ \psi^{L\beta}(x)\\ \psi^{S\alpha}(x)\\ \psi^{S\beta}(x)\end{pmatrix}.\label{4CWF}
\end{equation}
It can readily be checked that each of the four components satisfies the Klein-Gordon equation \eqref{KGeq}. It can also be shown that the density $\rho=\psi^\dag\psi$ is positive definite.
Moreover, the appearance of the Pauli spin matrices \eqref{PauliMat} in the Dirac $\boldsymbol{\alpha}$ matrix \eqref{DiracMat} reveals that the Dirac equation \eqref{fp-DEQ} is a relativistic wave equation for spin-$\frac{1}{2}$ particles, such that the components of the wave function \eqref{4CWF} can be labeled by the $\alpha$ and $\beta$ spins.

As a matter of fact, the Dirac equation \eqref{fp-DEQ} can directly be obtained from the Klein-Gordon equation \eqref{KGeq} by decomposing the latter into two coupled first-order equations.
Following Kramers\cite{KramersBook,Karwowski2017}, this can proceed by rewriting Eq. \eqref{KGeq} as
 \begin{equation}
 (i\hbar\frac{\partial}{\partial t}-c\boldsymbol{\sigma}\cdot\boldsymbol{p})(i\hbar\frac{\partial}{\partial t}+c\boldsymbol{\sigma}\cdot\boldsymbol{p})\psi_{\mathrm{KG}}(x)=m^2c^4\psi_{\mathrm{KG}}(x),
 \end{equation}
where use of the identity $\boldsymbol{p}^2=(\boldsymbol{\sigma}\cdot\boldsymbol{p})^2$ has been made. Further replacing $\psi_{\mathrm{KG}}$ with spinor $\psi^l$ and
$(i\hbar\frac{\partial}{\partial t}+c\boldsymbol{\sigma}\cdot\boldsymbol{p})\psi^l$ with $mc^2\psi^r$, we obtain
\begin{equation}
i\hbar\frac{\partial}{\partial t}\psi_{\mathrm{W}}(x)=D_0^{\mathrm{W}}\psi_{\mathrm{W}}(x),\label{Weyl}
\end{equation}
where
\begin{equation}
D_0^{\mathrm{W}}=\begin{pmatrix}c\boldsymbol{\sigma}\cdot\boldsymbol{p}&mc^2\\ mc^2&-c\boldsymbol{\sigma}\cdot\boldsymbol{p}\end{pmatrix},\quad \psi_{\mathrm{W}}(x)=\begin{pmatrix}\psi^r(x)\\ \psi^l(x)\end{pmatrix}.
\end{equation}
Eq. \eqref{Weyl} is known as the Dirac equation in the Weyl representation. By further carrying out the following unitary transformation,
\begin{equation}
U_{\mathrm{W}}=\frac{1}{\sqrt{2}}\begin{pmatrix} 1&1\\ 1&-1\end{pmatrix}=U_{\mathrm{W}}^{-1},
\end{equation}
the Dirac equation in the standard representation \eqref{fp-DEQ} can be recovered. At this moment it is worthy mentioning that,
although electron spin appears naturally in the Dirac equation, it is not a relativistic quantity, since it appears also in the
L{\'e}vy-Leblond equation\cite{LLeq}
\begin{equation}
i\hbar\frac{\partial}{\partial t}\begin{pmatrix} 1&0\\ 0&0\end{pmatrix}\begin{pmatrix}\psi^L(x)\\\phi^L(x)\end{pmatrix}
=\begin{pmatrix}0& \boldsymbol{\sigma}\cdot\boldsymbol{p}\\\boldsymbol{\sigma}\cdot\boldsymbol{p}&-2m\end{pmatrix}\begin{pmatrix}\psi^L(x)\\\phi^L(x)\end{pmatrix},
\end{equation}
which is just the nonrelativistic limit (nrl)\cite{DPT2} of the Dirac equation \eqref{fp-DEQ} and can be introduced \emph{a priori} by means of a spinor representation of
the (nonrelativistic) Galilei group.

\begin{figure}
\includegraphics[width=0.4\textwidth]{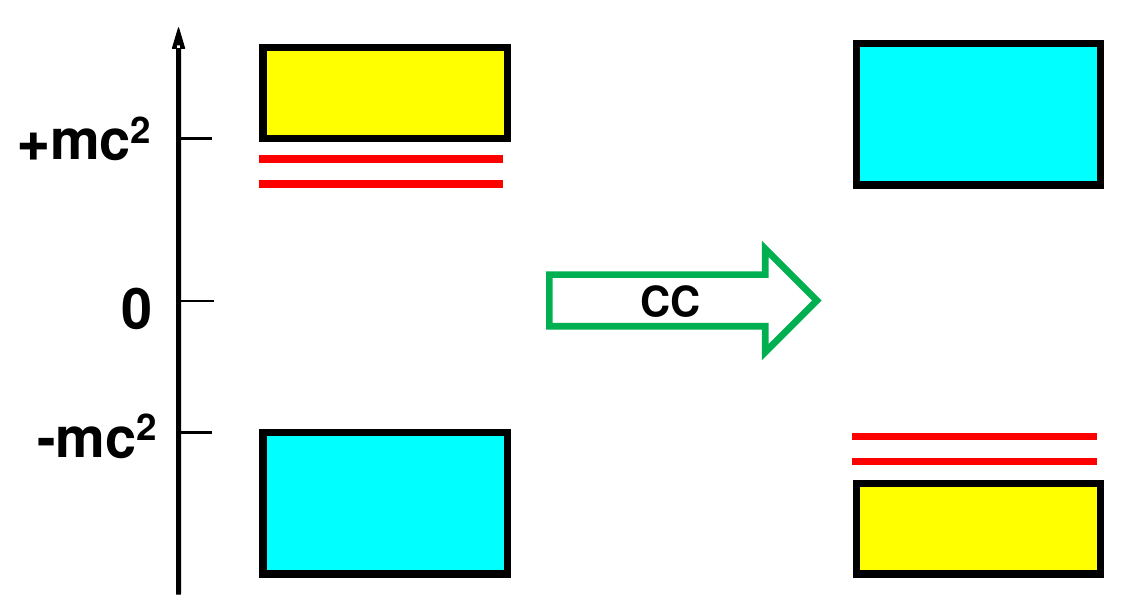}\\
\caption{Energy spectra of the electron (left) and positron (right) Dirac equations related by charge conjugation (CC).}
\label{DiracSpectrum}
\end{figure}

For an electron ($q=-e$) moving in an external electromagnetic field characterized by the vector potential $\boldsymbol{A}_{ext}$ and scalar potential $\phi_{ext}$,
the following minimal coupling relations
\begin{equation}
\boldsymbol{p}\rightarrow \boldsymbol{\pi}=\boldsymbol{p}-q\boldsymbol{A}_{ext},\quad i\hbar\frac{\partial}{\partial t}\rightarrow i\hbar\frac{\partial}{\partial t}-q\phi_{ext}
\end{equation}
for electromagnetic interaction can be invoked, so as to obtain
\begin{align}
&i\hbar\frac{\partial}{\partial t}\psi(x)=D\psi(x),\label{TD-DEQ}\\
&D=c\boldsymbol{\alpha}\cdot(\boldsymbol{p}-q\boldsymbol{A}_{ext})+\beta mc^2+q\phi_{ext}.\label{hDop}
\end{align}
If the external field is static, suffice it to consider the following eigenvalue problem
\begin{align}
D\psi_p(\boldsymbol{r})=\varepsilon_p\psi_p(\boldsymbol{r}),\label{DEQ}
\end{align}
which has three branches of solutions if the external field arises from a net 
positive charge distribution: positive-energy continuum, discrete positive-energy bound states and negative-energy continuum, as illustrated
by the left panel of Fig. \ref{DiracSpectrum}. The gap ($\Delta E$) between the lowest positive-energy level $\epsilon_{1s}$
and the top edge ($-mc^2$) of the negative-energy continuum can be calculated as
\begin{eqnarray}
\Delta E&=&\epsilon_{1s}-(-mc^2)=mc^2\sqrt{1-(Z/c)^2}+mc^2=2mc^2f_Z,\nonumber\\
\frac{1}{2}&<& f_Z=\frac{1}{2}(1+\sqrt{1-(Z/c)^2}) <1,
\end{eqnarray}
where $\epsilon_{1s}$ is the ground state energy of the Dirac equation for a one-electron atom of nuclear charge $Z$.
It is seen that the gap is indeed very large (e.g., $f_Z\approx 0.91$ for Hg$^{79+}$). It is not much changed
for many-electron systems ($f_Z\approx 0.92$ for Hg).

The existence of an empty negative-energy continuum was extremely troublesome in the early days of relativistic quantum mechanics, for it implied that
no atom would be stable! For instance, it can be estimated\cite{greiner1990relativistic} that, in the presence of a radiative field (which always exists
in reality), the electron in the ground state of the hydrogen atom
can fall down to the top of the negative-energy continuum in less than one nanosecond,
and it can even trigger a radiation catastrophe via continuous radiative transitions. To resolve this apparent untruth, Dirac\cite{DiracSea} proposed in 1930
that all states of negative energy be filled, such that transitions of electrons to negative-energy states
are forbidden by virtue of the Pauli exclusion principle. Since a hole left by exciting an electron from the filled
negative-energy sea has a positive energy and the same mass but opposite charge as the leaving electron,
it was interpreted by Dirac in 1931\cite{DiracPositron} as an anti-electron (positive electron/positron). Although highly controversial, this
bold prediction was confirmed by experiment just one year later\cite{AndersonPositron}. Notwithstanding such a big triumph, Dirac's hole theory
has a number of defects\cite{PCCPNES}:
\begin{enumerate}[(1)]
\item It is asymmetric with respect to electrons and positrons.
\item It characterizes a positron as a virtual hole rather than a real
particle.
\item It involves an infinite negative electric charge filling the
whole space even if only one electron is under consideration.
\item It has to assume that the negative-energy electrons do not
generate any potential acting on the positive-energy electrons. Otherwise, no nuclear charge could generate an enough attraction to
compensate the infinitely repulsive potential.
\item It does not, in a strict sense, explain the stability of a
positive-energy electron: being infinitely large, the sea can
always accept infinitely many electrons. In other words, the
Pauli exclusion principle does not really hold for a system of
an infinite number of fermions.
\item It does not apply to spin-0 particles (described by the Klein-Gordon equation \eqref{KGeq}) which do not
satisfy an exclusion principle.
\end{enumerate}

The above problems associated with the negative-energy continuum drove the pioneers of quantum mechanics to formulate a quantum field theory for electrodynamics (QED)
through a particular second quantization of the Dirac matter field and electromagnetic field, where all particles are of positive energy
(see Ref. \citenum{HistoryQED} for a historical review of the early days of QED).
To see how this can be achieved, we first take a look at the charge-conjugation transformation
\begin{equation}
\hat{C}=\mathbf{C}\beta\hat{K}_0,\quad \mathbf{C}=-i\alpha_y \label{CCsymm}
\end{equation}
of the Dirac equation \eqref{DEQ}: taking the complex conjugate ($\hat{K}_0$) followed by multiplying $\mathbf{C}\beta$
from the left leads to
\begin{equation}
[ c\boldsymbol{\alpha} \cdot(\boldsymbol{p} + q\boldsymbol{A}_{ext} ) +  \beta mc^2 - q \phi_{ext} ]\hat{C}\psi_p(\boldsymbol{r})=\hat{C}\psi_p(\boldsymbol{r})\epsilon_p^C,\quad \epsilon_p^C=-\epsilon_p.\label{PositronDEQ}
\end{equation}
The manipulation is facilitated by making use of the following identities
\begin{equation}
\mathbf{C}^\dag= \mathbf{C}^T = \mathbf{C}^{-1} = -\mathbf{C},\quad
\mathbf{C} \beta= -\beta\mathbf{C},\quad \mathbf{C}\boldsymbol{\alpha}^* = -\boldsymbol{\alpha}\mathbf{C},\quad \mathbf{C}^\dag \boldsymbol{\alpha}\mathbf{C}=-\boldsymbol{\alpha}^T.
\end{equation}
To unify the notion, we define
\begin{subequations}\label{PsiC-def}
\begin{equation}
\psi_{\tilde p}^C(\boldsymbol{x})=\psi_{\tilde p}^C(\boldsymbol{r})e^{-i\epsilon_{\tilde p}t}=\hat{C}\psi_p(x),\quad \psi^C_{\tilde p}(\boldsymbol{r})=\mathbf{C}\beta \psi_p^*(\boldsymbol{r}),\quad \epsilon_{\tilde p}=-\epsilon_p<0,\label{CC-4CWF-}
\end{equation}
\begin{equation}
\psi_{p}^C(\boldsymbol{x})=\psi_{p}^C(\boldsymbol{r})e^{-i\epsilon_{p}t}=\hat{C}\psi_{\tilde p}(x),\quad \psi^C_{p}(\boldsymbol{r})=\mathbf{C}\beta \psi_{\tilde p}^*(\boldsymbol{r}),\quad \epsilon_{p}=-\epsilon_{\tilde p}>0.\label{CC-4CWF+}
\end{equation}
\end{subequations}
That is, apart from its apparent action, charge conjugation will also interchange the indices of the argument as $p\leftrightarrow\tilde{p}$, such that
$\epsilon_p>0$ and $\epsilon_{\tilde p}<0$ always hold. Another example is $\hat{C}[a_p\psi_p]=\mathbf{C}\beta(a_p\psi_p)^{\dag T}=\mathbf{C}\beta(a^p\psi_p^*)=a^{\tilde p}\psi_{\tilde p}^C$.

It is clear that, for the same time-independent external field $(\phi_{ext}(\boldsymbol{r}), \boldsymbol{A}_{ext}(\boldsymbol{r}))$, if $\psi_p(x)=\psi_p(\boldsymbol{r})e^{-i\epsilon_p t}$ is a
stationary state of the Dirac equation for an electron ($q=-1$) of positive energy $\epsilon_p$,
$\psi_{\tilde p}^C(x)=\hat{C}\psi_p(x)=\psi_{\tilde p}^C(\boldsymbol{r})e^{-i\epsilon_{\tilde p} t}$ will then be a
stationary state of the Dirac equation
for a positron ($q=+1$) of negative energy $\epsilon_{\tilde p}=-\epsilon_p$. Likewise,
if $\psi_{\tilde p}(x)=\psi_{\tilde p}(\boldsymbol{r})e^{-i\epsilon_{\tilde p} t}$ is an electronic negative-energy state (NES),
$\psi_p^C(x)=\hat{C}\psi_{\tilde p}(x)=\psi_p^C(\boldsymbol{r})e^{-i\epsilon_p t}$ will be a
positronic positive-energy state (PES; $\epsilon_p=-\epsilon_{\tilde p}$;
cf. the right panel of Fig. \ref{DiracSpectrum}). Note in particular that
the probability density of a negative-energy electron $\psi_{\tilde p}(\boldsymbol{r})e^{-i|\epsilon_{\tilde p}|(-t)}$
is indistinguishable from that of a positive-energy positron $\psi_p^C(\boldsymbol{r})e^{-i|\epsilon_{\tilde p}| t}$, i.e.,
$|\psi_p^C(\boldsymbol{r})|^2=|\psi_{\tilde p}(\boldsymbol{r})|^2$. As such, a negative-energy electron propagating backward in time can be regarded as the mirror image of a positive-energy positron propagating
forward in time. Therefore, the Dirac matter field should be quantized as
\begin{subequations}\label{Fieldbdef}
\begin{equation}
\hat{\phi}(x)=b_p\psi_{p}(x)+b^{\tilde p}\psi_{\tilde{p}}(x),\label{Fieldb}
\end{equation}
\begin{equation}
b_p|0\rangle=b_{\tilde p}|0\rangle=0,\quad b^p=b_p^\dag,\quad b^{\tilde p}=b_{\tilde p}^\dag,\quad \epsilon_p>0,
\quad \epsilon_{\tilde p}<0, \label{b-op}
\end{equation}
\end{subequations}
in the interaction picture and the Einstein summation convention
over repeated indices, in order for the field to comprise only of positive-energy particles:
$b_p$ annihilates an electron of positive energy $\epsilon_p$, whereas
$b^{\tilde p}$ creates a positron of positive energy $|\epsilon_{\tilde p}|= -\epsilon_{\tilde p}$. Since
any operator must be expanded in a complete (orthonormal) basis spanned by the PES and NES
of the same Dirac equation \eqref{DEQ}, the amplitude companying $b^{\tilde p}$ can only be the electronic NES $\psi_{\tilde{p}}(\boldsymbol{r})e^{-i\epsilon_{\tilde p} t}$
instead of the corresponding positronic PES $\psi_p^C(\boldsymbol{r})e^{-i|\epsilon_{\tilde p}|t}$ (NB: in the presence of an external field,
$\{\psi_p^C(\boldsymbol{r})\}$ are even not orthogonal to the electronic PES $\{\psi_{q}(\boldsymbol{r})\}$, i.e., the inner products $\langle \psi_p^C|\psi_{q}\rangle$ are generally nonzero).
On the other hand, charge conservation dictates that the operator $b^{\tilde p}$ (instead of $b_{\tilde p}$) must accompany $b_p$.
Both $b_p$ and $b^{\tilde p}$ increase the charge of a state by one unit; $b_p$ does this by destroying an electron whereas $b^{\tilde p}$
does this by creating a positron. Thus the field operator $\hat{\phi}(x)$ always increases one unit of charge. Similarly,
the field operator $\hat{\phi}^\dag(x)$ always decreases one unit of charge. Therefore, the operator $\hat{\phi}^\dag(x)\hat{\phi}(x)$ conserves the charge.
Had $b_{\tilde p}$ been chosen to accompany $b_p$, the operator $\hat{\phi}^\dag(x)\hat{\phi}(x)$ would not conserve the charge:
it would include terms like $b^pb_{\tilde p}$ and $b^{\tilde p}b_p$ which decrease and increase
two units of charge, respectively.

The particular form \eqref{Fieldbdef} for the quantized Dirac matter field is the very first cornerstone of QED. It
can actually be rewritten as\cite{PCCPNES}
\begin{subequations}\label{Fieldadef}
\begin{equation}
\hat{\phi}(x)=a_p\psi_{p}(x)+a_{\tilde p}\psi_{\tilde{p}}(x),\label{Fielda}
\end{equation}
\begin{equation}
a_p|0_{e^-};N_{e^-}\rangle=a^{\tilde p}|0_{e^-};N_{e^-}\rangle=0,\quad a^p=a_p^\dag,\quad a^{\tilde p}=a_{\tilde p}^\dag,\quad \epsilon_p>0,
\quad \epsilon_{\tilde p}<0, \label{a-op}
\end{equation}
\end{subequations}
by replacing the genuine vacuum $|0\rangle$ with the physical vacuum $|0_{e^-};N_{e^-}\rangle$ consisting of zero positive-energy electrons and $N_{e^-}$ ($\rightarrow\infty$) negative-energy electrons.
That is, the particle-hole picture
\begin{equation}
b_p=a_p,\quad b^p=a^p,\quad b_{\tilde p}=a^{\tilde p},\quad b^{\tilde p}=a_{\tilde p},\quad \epsilon_p>0,\quad \epsilon_{\tilde p}<0 \label{a2b-op}
\end{equation}
implied in Eq. \eqref{Fieldb} is merely a mathematical operation and is only convenient for pictorial interpretation in terms of diagrams,
where the expression \eqref{Fielda} is more convenient for algebraic manipulations\cite{PCCPNES}. At first glance,
we have just gone back to the filled Dirac picture, such that the aforementioned problems associated with Dirac's hole theory
would arise again. However, the picture can be reversed: it is perfectly legitimate
to quantize the Dirac matter field in terms of the solutions of the positron Dirac equation \eqref{PositronDEQ}:
\begin{subequations}\label{Field-Pbdef}
\begin{equation}
\hat{\phi}^C(x)=d_p\psi_{p}^C(x)+d^{\tilde p}\psi_{\tilde{p}}^C(x),\label{Field-Pb}
\end{equation}
\begin{equation}
d_{p}|0\rangle=d_{\tilde p}|0\rangle=0,\quad d^p=d_p^\dag,\quad d^{\tilde p}=d_{\tilde p}^\dag,\quad \epsilon_p>0,
\quad \epsilon_{\tilde p}<0, \label{d-op}
\end{equation}
\end{subequations}
where $d_p$ annihilates a positron of positive energy $\epsilon_p$, whereas $d^{\tilde p}$ creates an electron of positive energy $|\epsilon_{\tilde p}|$.
The charge-conjugation transformation of $\hat{\phi}(x)$ \eqref{Fieldbdef} leads to\cite{eQEDBook2017}\footnote{The charge-conjugation transformation \eqref{CTfield}
of a field operator is bound to the particle-hole picture and hence should not be applied to expressions \eqref{Fielda} and \eqref{Field-Pc}.}
\begin{eqnarray}
\hat{\phi}^C(x)&=&\mathbf{C}\beta\hat{\phi}^{\dag T}(x)\label{CTfield}\\
&=&\mathbf{C}\beta[b^p\psi_p^*(x)]+\mathbf{C}\beta[b_{\tilde p}\psi_{\tilde p}^*(x)]\\
&=&b^{\tilde p}\psi_{\tilde p}^C(x)+b_p\psi_p^C(x),\quad \mbox{s.t. } \epsilon_p>0, \epsilon_{\tilde p}<0.\label{junk2}
\end{eqnarray}
By comparing Eq. \eqref{junk2} with Eq. \eqref{Field-Pb} we obtain
\begin{equation}
d_p=b_p,\quad d^{\tilde p}=b^{\tilde p}, \quad \epsilon_p>0, \quad \epsilon_{\tilde p}<0.
\end{equation}
That is, the $d$ and $b$ types of annihilation and creation processes are the same, although their carriers
are different (positrons vs electrons). This is more transparent\cite{eQEDBook2017} for the case of free particles for which Eqs. \eqref{DEQ} and \eqref{PositronDEQ}
are identical (i.e., $\psi_p^C(x)=\psi_p(x)$ and $\psi_{\tilde p}^C(x)=\psi_{\tilde p}(x)$), such that it is immaterial to interpret which set of the PES and NES
as electrons or positrons. More over, just like Eq. \eqref{Fieldadef},
Eq. \eqref{Field-Pbdef} can be rewritten as
\begin{subequations}\label{Field-Pcdef}
\begin{equation}
\hat{\phi}^C(x)=c_p\psi_{p}^C(x)+c_{\tilde p}\psi_{\tilde{p}}^C(x),\label{Field-Pc}
\end{equation}
\begin{equation}
c_{p}|0_{e^+};N_{e^+}\rangle=c^{\tilde p}|0_{e^+};N_{e^+}\rangle=0,\quad c^p=c_p^\dag,\quad c^{\tilde p}=c_{\tilde p}^\dag,\quad \epsilon_p>0,
\quad \epsilon_{\tilde p}<0. \label{c-op}
\end{equation}
\end{subequations}
Now the vacuum is $|0_{e^+};N_{e^+}\rangle$ in lieu of the original $|0\rangle$.
Since the two types of (second) quantization of the same Dirac matter field are equivalent, they can simply be averaged with an equal weight.
To show how this can be done, let us look at the four-current operators for electrons and positrons:
\begin{eqnarray}
\hat{j}^\mu_{e^-}(x)&=&-e\hat{\phi}^\dag(x)c\alpha^{\mu}\hat{\phi}(x), \quad \alpha^{\mu}=(c^{-1},\boldsymbol{\alpha}),\quad e=+1,\\
&=&-e\{\hat{\phi}^\dag(x)c\alpha^{\mu}\hat{\phi}(x)\}-e\langle 0;N_{e^-}|\hat{\phi}^\dag(x)c\alpha^{\mu}\hat{\phi}(x)|0;N_{e^-}\rangle\label{kunk1}\\
&=&-e\{\hat{\phi}^\dag(x)c\alpha^{\mu}\hat{\phi}(x)\}-e\psi_{\tilde p}^\dag(\boldsymbol{r}) c\alpha^{\mu}\psi_{\tilde p}(\boldsymbol{r})\label{kunk1b},\\
\hat{j}^\mu_{e^+}(x)&=&e\hat{\phi}^{C\dag}(x)c\alpha^{\mu}\hat{\phi}^C(x)\label{j+Cdef}\\
&=&e\{\hat{\phi}^{C\dag}(x)c\alpha^{\mu}\hat{\phi}^C(x)\}+e\langle 0;N_{e^+}|\hat{\phi}^{C\dag}(x)c\alpha^{\mu}\hat{\phi}^C(x)|0;N_{e^+}\rangle\label{kunk2}\\
&=&e\{\hat{\phi}^{C\dag}(x)c\alpha^{\mu}\hat{\phi}^C(x)\}+e\psi_{\tilde p}^{C\dag}(\boldsymbol{r})c\alpha^{\mu}\psi_{\tilde p}^C(\boldsymbol{r}),\label{junk3}
\end{eqnarray}
where the first terms of Eqs. \eqref{kunk1}/\eqref{kunk1b} and \eqref{kunk2}/\eqref{junk3} are normal ordered with respect to $|0;N_{e^-}\rangle$ and $|0;N_{e^+}\rangle$,
respectively. By means of the relation \eqref{CTfield}, the first term of Eq. \eqref{junk3} can be written as
\begin{eqnarray}
e\{\hat{\phi}^{C\dag}(x)c\alpha^{\mu}\hat{\phi}^C(x)\}&=&ec\{\hat{\phi}^T\beta\mathbf{C}^\dag\alpha^{\mu}\mathbf{C}\beta\hat{\phi}^{\dag T}(x)\} \label{phiCdef}\\
&=&ec\{\hat{\phi}^T_\gamma(x)(\alpha^\mu)^T_{\gamma\rho}\hat{\phi}^{\dag T}_{\rho}(x)\}\\
&=&e\{c\alpha^{\mu}\hat{\phi}(x)\hat{\phi}^\dag(x)\}\label{lunk2}\\
&=&-e\{\hat{\phi}^\dag(x)c\alpha^{\mu}\hat{\phi}(x)\},
\end{eqnarray}
where the normal ordering is now taken with respect to $|0;N_{e^-}\rangle$.
Likewise, the second term of Eq. \eqref{junk3} can be written as
\begin{eqnarray}
e\psi_{\tilde p}^{C\dag}(\boldsymbol{r})c\alpha^{\mu}\psi_{\tilde p}^C(\boldsymbol{r})=e\psi_p^\dag(\boldsymbol{r}) c\alpha^{\mu}\psi_p(\boldsymbol{r})=e\langle 0;N_{e^-}|c\alpha^{\mu}\hat{\phi}(x)\hat{\phi}^\dag(x)|0;N_{e^-}\rangle.
\end{eqnarray}
Therefore, $\hat{j}^\mu_{e^+}(x)$ \eqref{j+Cdef} can be written as
\begin{eqnarray}
\hat{j}^\mu_{e^+}(x)&=&ec\alpha^{\mu}\hat{\phi}(x)\hat{\phi}^\dag(x)\label{j+def}\\
&=&-e\{\hat{\phi}^\dag(x)c\alpha^{\mu}\hat{\phi}(x)\}+e\langle 0;N_{e^-}|c\alpha^{\mu}\hat{\phi}(x)\hat{\phi}^\dag(x)|0;N_{e^-}\rangle.
\end{eqnarray}
The four-current operator averaged over electrons and positrons then reads
\begin{eqnarray}
\hat{j}^\mu(x)&=&\frac{1}{2}(\hat{j}^\mu_{e^-}(x)+\hat{j}^\mu_{e^+}(x))\\
&=&-\frac{1}{2}e[\hat{\phi}^\dag(x), c\alpha^{\mu}\hat{\phi}]=-\frac{1}{2}ec\alpha^\mu_{\gamma\rho}[\hat{\phi}_{\gamma}^\dag(x),\hat{\phi}_{\rho}(x)]\label{4-current1}\\
&=&-e\{\hat{\phi}^\dag(x)c\alpha^{\mu}\hat{\phi}\}+j^\mu_{vp}(\boldsymbol{r}),\label{4-current2}\\
j^\mu_{vp}(\boldsymbol{r})&=&-e\langle \mathrm{vac}|\frac{1}{2}[\hat{\phi}^\dag(x), c\alpha^{\mu}\hat{\phi}(x)]|\mathrm{vac}\rangle\label{jvacdef}\\
&=&-\frac{1}{2}e [\psi_{\tilde p}^\dag(\boldsymbol{r})c\alpha^\mu \psi_{\tilde p}(\boldsymbol{r})-\psi_p^\dag(\boldsymbol{r})c\alpha^\mu\psi_p(\boldsymbol{r})].\label{vacCurrent}
\end{eqnarray}
Note in passing that the vacuum $|\mathrm{vac}\rangle$ in Eqs. \eqref{4-current2} and \eqref{jvacdef}
can either be $|0\rangle$ along with the definition \eqref{Fieldbdef}
or $|0_{e^-};N_{e^-}\rangle$ along with the definition \eqref{Fieldadef}.
The zero component of $j^\mu_{vp}(\boldsymbol{r})$ \eqref{vacCurrent}, i.e., the vacuum density $\rho_{vp}(\boldsymbol{r})$, reads
\begin{eqnarray}
\rho_{vp}(\boldsymbol{r})&=&-e\langle \mathrm{vac}|\frac{1}{2}[\hat{\phi}^\dag(x), \hat{\phi}(x)]|\mathrm{vac}\rangle\label{rhovapdef}\\
&=&-\frac{1}{2}e [n_-(\boldsymbol{r})-n_+(\boldsymbol{r})],\label{SDenvp}\\
n_+(\boldsymbol{r})&=&\psi_p^\dag(\boldsymbol{r})\psi_p(\boldsymbol{r})=\psi_{\tilde p}^{C\dag}(\boldsymbol{r})\psi_{\tilde p}^C(\boldsymbol{r}),\label{n+def}\\
 n_-(\boldsymbol{r})&=&\psi_{\tilde p}^\dag(\boldsymbol{r})\psi_{\tilde p}(\boldsymbol{r})=\psi_p^{C\dag}(\boldsymbol{r})\psi^C_p(\boldsymbol{r}),\label{n-def}
\end{eqnarray}
where $n_+(\boldsymbol{r})$ and $n_-(\boldsymbol{r})$ are the number densities of the PES and NES of the electron Dirac equation \eqref{DEQ}, respectively.
By virtue of the identity $n_+ + n_- = {\bar n}_+ + {\bar n}_- = 2{\bar n}_-$, with ${\bar n}_+$ and ${\bar n}_-$ ($={\bar n}_+$)
being the free-particle number densities, we have
\begin{eqnarray}
\rho_{vp}(\boldsymbol{r})&=&-e [n_-(\boldsymbol{r}) - {\bar n}_-(\boldsymbol{r})],\label{N-+}
\end{eqnarray}
which is clearly the charge polarization of the vacuum. Moreover, Eq. \eqref{SDenvp} reveals that the NES of the electron Dirac equation \eqref{DEQ} are all occupied by electrons $e^-$ with charge -1
(i.e., filled Dirac sea of electrons), whereas the PES by positrons $e^+$ with charge +1. As shown above,
the latter arises actually from the filled Dirac sea of positrons (i.e., $|0_{e^+};N_{e^+}\rangle$), as a direct consequence of charge conjugation.
Therefore, the genuine vacuum $|0\rangle$ can be viewed\cite{IJQCeQED} as the superposition of $|0_{e^-}; N_{e^-}\rangle$ and $|0_{e^+};N_{e^+}\rangle$: the electrons and positrons annihilate each other spontaneously, so as to leave an empty vacuum. In other words, the original hole theory of Dirac\cite{DiracSea}
for relativistic electrons should be generalized to ``charge-conjugated hole theory'' or simply ``extended hole theory''.
This feature is incorporated automatically into the symmetrized four-current operator \eqref{4-current1}, introduced first by Schwinger
in 1951\cite{Schwinger1951}.
Given its great importance, the expression \eqref{4-current1} should be viewed as another cornerstone of QED. Note in passing that,
in the free-particle (fp) representation, $j^\mu_{vp}(\boldsymbol{r})$ \eqref{jvacdef} vanishes pointwise, thereby leading to
$\hat{j}_{fp}^\mu(x)=-e\{\hat{\phi}^\dag(x)c\alpha^{\mu}\hat{\phi}\}$.

Last but not least, it is worthy mentioning that the commutator form of contraction \eqref{rhovapdef} is just a special case of the equal-time contraction (ETC)\cite{Schwinger1951} of fermion operators
\begin{eqnarray}
\acontraction[0.5ex]{}{A}{(t)}{B}A(t)B(t)&=&\langle \mathrm{vac}| T[A(t)B(t)]|\mathrm{vac}\rangle\\
&\triangleq&\frac{1}{2}\langle\mathrm{vac}| T[A(t)B(t^\prime)] |\mathrm{vac}\rangle|_{t^\prime-t\rightarrow 0^{\pm}} \\
&=&\langle \mathrm{vac}|\frac{1}{2}[A(t), B(t)]|\mathrm{vac}\rangle,\label{ETC}
\end{eqnarray}
which is symmetric in time. That is, the two expressions $A(t)B(t)$ and $-B(t)A(t)$ obtained by letting $t^\prime$ approach $t$ from the past and future
are both considered and averaged here. Eq. \eqref{ETC} is fundamentally different from the following ETC
\begin{eqnarray}
\bcontraction[0.5ex]{}{A}{(t)}{B}A(t)B(t)=\lim_{\eta\rightarrow 0^+}\langle 0;\tilde{0}| T[A(t)B(t+\eta)] |0;\tilde{0}\rangle,\label{NRETC}
\end{eqnarray}
which is asymmetric in time and holds only in the nrl.
Note that the ETC \eqref{ETC} is only implicit in the Feynman fermion propagator\cite{IJQCeQED}, such that its importance is often overlooked in the literature.
Instead, we should regard it as an essential ingredient to distinguish relativistic from nonrelativistic quantum mechanics.
As a time-independent analog of the ETC \eqref{ETC}, the charge-conjugated contraction (CCC) of fermion operators was also introduced\cite{eQED},
\begin{eqnarray}
\acontraction[0.5ex]{}{a^p}{}{a_q}a^pa_q&=&\langle 0;N_{e^-}|\frac{1}{2}[a^p, a_q]|0;N_{e^-}\rangle,\quad p, q \in \mbox{PES, NES}\label{CCC0}\\
&=& \frac{1}{2}\langle 0;N_{e^-}|a^{\tilde{p}}a_{\tilde{q}}|0;N_{e^-}\rangle|_{\epsilon_{\tilde{p}}<0, \epsilon_{\tilde{q}}<0}
   -\frac{1}{2}\langle 0;N_{e^-}|a_qa^p                    |0;N_{e^-}\rangle|_{\epsilon_p>0, \epsilon_q>0}\label{CCsym}\\
&=&-\frac{1}{2}\delta^p_q \sgn(\epsilon_q), \quad p, q \in \mbox{PES, NES},\label{CCC}
\end{eqnarray}
which distinguishes from the standard contraction
\begin{eqnarray}
\bcontraction[0.5ex]{}{a^p}{}{a_q}a^pa_q&=&\langle 0;N_{e^-}|a^pa_q|0;N_{e^-}\rangle,\quad p, q \in \mbox{PES, NES}\\
&=&\delta^{\tilde{p}}_{\tilde{q}}n_{\tilde{q}}.\label{NRCCC2}
\end{eqnarray}
Although the introduction of CCC \eqref{CCC} looks very trivial, it is a key ingredient in a time-independent Fock space formulation of relativistic quantum mechanics. In particular,
it allows to construct\cite{eQED,IJQCeQED} an effective QED (eQED) Hamiltonian in a bottom-up fashion (i.e., without ever recourse to QED, a time-dependent perturbation theory). In contrast,
the standard contraction \eqref{NRCCC2}, the time-independent analog of Eq. \eqref{NRETC}, will result in wrong, nonrelativistic type of potential energy expressions even for relativistic operators.
As an illustration, we look at the number operator, which reads
\begin{eqnarray}
\hat{N}&=& a^p_p+a^{\tilde p}_{\tilde p}\\
&=&\{a^p_p\}+\{a^{\tilde p}_{\tilde p}\}+\langle 0;N_{e^-}|\frac{1}{2}[a^p, a_p]|0;N_{e^-}\rangle+\langle 0;N_{e^-}|\frac{1}{2}[a^{\tilde p}, a_{\tilde p}]|0;N_{e^-}\rangle\\
&=&\{a^p_p\}+\{a^{\tilde p}_{\tilde p}\} -\frac{1}{2}\delta^p_p+\frac{1}{2}\delta^{\tilde p}_{\tilde p}\\
&=&\{a^p_p\}+\{a^{\tilde p}_{\tilde p}\}\label{Number-a}
\end{eqnarray}
according to Eq. \eqref{CCC}, but reads
\begin{eqnarray}
\hat{N}=\{a^p_p\}+\{a^{\tilde p}_{\tilde p}\}+N_{e^-}\label{Number-c}
\end{eqnarray}
according to Eq. \eqref{NRCCC2}, with $N_{e^-}\rightarrow\infty$ in line with the filled Dirac sea. It can readily be checked that
the correct \eqref{Number-a} and incorrect \eqref{Number-c} results
can also be obtained by using the contractions \eqref{CCC} and \eqref{NRCCC2}, respectively, in terms of the $b$-operators \eqref{a2b-op}
 and the associated vacuum $|0\rangle$.

Having discussed pedagogically the basics of QED (including first quantization \eqref{TD-DEQ} of special relativity \eqref{Esquare}, second quantization of the Dirac field \eqref{Fieldbdef} or equivalently
\eqref{Fieldadef}, extended hole theory, symmetrized 4-current \eqref{4-current1}, equal-time contraction \eqref{ETC}, as well as
charge-conjugated contraction \eqref{CCC}), we just comment briefly on the applications of QED.
Undoubtedly, QED is the most accurate theory ever designed in physics. For instance, the anomalous magnetic
moment, $(g-2)/2$, of the electron has been determined to the 11th decimal place\cite{Electron-g}, which leads further to improved values for the
electron mass\cite{Electron-m} and the fine structure constant\cite{Electron-alpha}.
However, the situation is very different for bound states of many-electron systems\cite{BoundQED2017}, for which QED is computationally too expensive:
the more the electrons, the higher order of perturbation and hence the more Feynman diagrams are required to achieve high precision. Because of this,
relativistic QED has thus far been applied successfully only to single ions of at most 5 electrons (see Refs. \citenum{PyykkoChemRev2012} and \citenum{QEDeffects} for recent reviews).
As for molecular systems, only nonrelativistic QED has been applied to the lightest molecules (e.g., \ce{H2}\cite{H2QED2009}, \ce{D2}\cite{D2QED2010} and \ce{HD}\cite{HDQED2010}, etc.).
So the question is how to account for QED effects in heavy atoms and molecules.
To show relevance of this question, we just quote a few results here: (1) according to the rough estimates\cite{Dyall-Lamb-2001},
the leading-order QED (Lamb shift) effects can be as large as 1 kcal/mol in chemical processes involving heavy elements. (2)
According to the most recent and to date most accurate relativistic calculations\cite{SchwerdtfegerAuPRL2017} of the first ionization potential (IP) and electron affinity (EA) of the gold atom, QED effects are roughly the same as electron correlation beyond the gold standard CCSD(T) (coupled-cluster with singles and doubles and perturbative triples).
(3) As for core properties such as the K-edge electron spectra, QED effects become significant already for the third row of the periodic table\cite{CoreQED2017}. It is therefore clear that we do need a feasible
relativistic QED approach for the electronic structure and spectroscopies of heavy atoms and even molecules.
It is also clear that, at variance with the ``relativity-QED then correlation'' paradigm of QED,
we should think of something like ``first relativity then correlation and finally QED''\cite{IJQCrelH}.
Such effective QED (eQED) approaches\cite{eQED,IJQCeQED,PhysRep,np-eQED} do exist, which
will be discussed in Sec. \ref{SeceQED}. Before this, we need to know how to solve the time-independent
Dirac equation \eqref{DEQ} via a finite basis expansion (see Sec. \ref{SecDEQ}).  After having presented the eQED Hamiltonians in Sec. \ref{SeceQED}, we will discuss in Sec. \ref{SecNES}
the correlation problem of NES (or virtual positrons) as well as a relativistic theory of real positrons.
Sec. \ref{Sec2C} is devoted to a summary of no-pair relativistic Hamiltonians, whereas Sec. \ref{SecEc}
to the no-pair correlation problem.
The account will be closed with perspectives in Sec. \ref{Conclusion}.
\section{The matrix Dirac equation}\label{SecDEQ}
For brevity, consider first the Dirac equation for an electron moving in a local potential $V$
\begin{eqnarray}
\begin{pmatrix}V&c \boldsymbol{\sigma}\cdot \boldsymbol{p}
\cr c \boldsymbol{\sigma}\cdot \boldsymbol{p}&
V-2c^2\cr\end{pmatrix} \begin{pmatrix}\psi^L_p\cr
\psi_p^S\end{pmatrix}=\begin{pmatrix}\psi^L_p\cr
\psi_p^S\end{pmatrix}\epsilon_p,\label{effDEQ}
\end{eqnarray}
where the rest-mass energy $mc^2$ has been subtracted to align the energy scale to that of the Schr{\"o}dinger equation.
Early attempts\cite{Kim1967} to solve this equation in a basis expansion were plagued by the occurrence of matrix eigenvalues in
the forbidden region between the lowest positive-energy and the highest negative-energy operator eigenvalues.
This phenomenon is usually called variational collapse and is often
traced back to the lack of a lower-bound property
of the Dirac operator. Actually, the ``variational collapse''
is due to the fact that inappropriately chosen basis sets are unable to
describe the kinetic energy correctly and to guarantee the correct nrl\cite{SchwarzCPL}.
It can  be removed rigorously via the minimax principle\cite{TalmanMinimax,SereMinimax}, without the need to impose a lower-bound property
on the matrix representation of the Dirac equation \eqref{effDEQ}. On the practical side,
several prescriptions have been proposed to construct suitable basis sets, including restricted kinetic balance (RKB)\cite{RKB},
unrestricted kinetic balance (UKB)\cite{RKBUKB}
[NB: the acronyms RKB and UKB were first coined by Dyall and F{\ae}gri Jr.\cite{RKBname}], dual kinetic balance (DKB)\cite{DKB}
and inverse kinetic balance (IKB)\cite{IKB}. According to the thorough formal and numerical analyses\cite{IKB}, the following conclusions can be drawn:
\begin{enumerate}[(I)]
\item RKB is the least adequate condition for constructing the small-component spinor basis $\{f_{\mu}\}_{\mu=1}^{2N^L}$ directly from
the large-component set $\{g_{\mu}\}_{\mu=1}^{2N^L}$, viz.,
\begin{eqnarray}
f_{\mu}=\frac{\alpha}{2}\boldsymbol{\sigma}\cdot\boldsymbol{p}g_{\mu},\quad \alpha=c^{-1}, \quad\mu=1,\cdots, 2N^L. \label{RKB}
\end{eqnarray}
Why this is the case can best be understood in terms of the modified Dirac equation\cite{mDiracKutz,mDiracKen}
\begin{eqnarray}
D^M\psi^M_p &=& S^M \psi^M_p\epsilon_p,\label{MDEQ}\\
 {\cal T}_M&=&\begin{pmatrix}1&0\cr 0&\frac{\alpha}{2}\boldsymbol{\sigma}\cdot\boldsymbol{p}\end{pmatrix},\label{TRKB}\\
D^M&=&{\cal T}_M^{\dag}D{\cal T}_M=\begin{pmatrix}V&T\cr
T&\frac{\alpha^2}{4}\boldsymbol{\sigma}\cdot\boldsymbol{p}V\boldsymbol{\sigma}\cdot\boldsymbol{p}-T\end{pmatrix},\\
S^M&=&{\cal T}_M^{\dag}{\cal T}_M=\begin{pmatrix}1&0\cr 0&\frac{\alpha^2}{2}T\end{pmatrix},\\
\psi^M_p&=&{\cal T}_M^{-1}\psi_p=\begin{pmatrix}\psi^L_p\cr \phi^L_p\end{pmatrix},\\ \psi^S_p&=&\frac{\alpha}{2}\boldsymbol{\sigma}\cdot\boldsymbol{p}\phi^L_p.\label{PseudoLarge}
\end{eqnarray}
It has been proven\cite{LiuMP} that the large ($\psi_p^L$) and pseudo-large ($\phi_p^L$) components must be expanded in the same spinor basis $\{g_\mu\}$
in order to guarantee the correct nrl, a prerequisite to ensure that the energies of the PES are correct to $\mathcal{O}(c^{-2})$. Relation
\eqref{PseudoLarge} then implies immediately the RKB \eqref{RKB}. The expansion of $\psi_p$ in the RKB basis \eqref{RKB} and that of $\psi^M_p$ in the $\{g_{\mu}\}$ basis, i.e.,
\begin{eqnarray}
\psi_p&=&\begin{pmatrix}\psi_p^L\\ \psi_p^S\end{pmatrix}=
\begin{pmatrix}g_{\mu}A_{\mu p}\cr 0\end{pmatrix}+\begin{pmatrix}0\cr f_{\mu}B_{\mu p}\end{pmatrix},\label{RKBexpan}\\
\psi^M_p&=&\begin{pmatrix}\psi_p^L\\ \phi_p^L\end{pmatrix}=\begin{pmatrix}g_{\mu}A_{\mu p}\cr 0\end{pmatrix}+\begin{pmatrix}0\cr g_{\mu}B_{\mu p}\end{pmatrix},
\end{eqnarray}
give rise to the same matrix Dirac equation
\begin{eqnarray}
\begin{pmatrix}\mathbf{V} &\mathbf{T}\cr
\mathbf{T}&\frac{\alpha^2}{4}\mathbf{W}-\mathbf{T}\cr\end{pmatrix}\begin{pmatrix}\mathbf{A}_p\cr
\mathbf{B}_p\end{pmatrix}=\begin{pmatrix}\mathbf{S}&0\cr
0&\frac{\alpha^2}{2}\mathbf{T}\cr\end{pmatrix} \begin{pmatrix}\mathbf{A}_p\cr
\mathbf{B}_p\end{pmatrix}\epsilon_p, \label{DEQMat}
\end{eqnarray}
where the individual matrices are all of dimension $2N^L$, with the elements being
\begin{eqnarray}
V_{\mu\nu}&=&\langle g_{\mu}|V|g_{\nu}\rangle,\quad
T_{\mu\nu}=\langle g_{\mu}|\frac{p^2}{2}|g_{\nu}\rangle,\nonumber\\
W_{\mu\nu}&=&
\langle g_{\mu}|\boldsymbol{\sigma}\cdot\boldsymbol{p}V\boldsymbol{\sigma}\cdot\boldsymbol{p}|g_{\nu}\rangle,\quad
S_{\mu\nu}=\langle g_{\mu}|g_{\nu}\rangle.
\end{eqnarray}
Eq.~\eqref{DEQMat} is therefore of dimension $4N^L$ with
$2N^L$ PES and $2N^L$ NES, which are separated by ca. $2mc^2\approx 1$ MeV.
When solving the equation \eqref{DEQMat} iteratively, the energetically lowest PES are chosen to be occupied
in each iteration cycle, so as to avoid variational collapse. While the rotations
between the occupied and unoccupied PES lower the total energy, those between the occupied PES and
unoccupied NES raise the total energy, to a much lesser extent though.

The following points concerning RKB still deserve to be highlighted. (a) The RKB condition does not provide full variational safety, because
the NES are in error of $\mathcal{O}(c^0)$\cite{IKB}.
Depending very much on the construction of the large-component basis, some bounds failures (or prolapse\cite{Prolapse}) of $O(c^{-4})$ may occur.
Nevertheless, such bounds failures will diminish when approaching to the basis set limit, at a rate that is
not much different from the nonrelativistic counterpart\cite{kutzelnigg2007completeness}. (b) It turns out that
the use of spherical Gaussians with principal quantum number $n$ larger than the angular momentum $l$ plus one leads to terrible variational collapse\cite{IKB},
although such functions are valid in the nonrelativistic case. Therefore, the use of spherical Gaussians subject to the restriction
$n=l+1$ (i.e., $1s$, $2p$, $3d$, $4f$, $5g$, etc.) is not merely a matter of economy but also a must.

\item IKB is the charge-conjugated version of RKB. It guarantees the correct nrl for the NES instead of the PES. Because of this, it requires
basis functions that are very different from the standard ones and is therefore only of conceptual interest rather than of practical usage.

\item DKB combines the good of both RKB and IKB and even provides full variational safety\cite{IKB}. However, such
an advantage is largely offset by its complicated nature and doubled number of integrals compared to RKB.
It is therefore recommended only for calculations of tiny quantities (e.g., QED and parity non-conserving effects), where
the complexity of DKB is only minor compared to the high precision to be achieved.

\item UKB is not uniquely defined. A scalar UKB basis does not transform as the basis of irreducible representations of double point groups or of time-reversal symmetry, a not serious problem though. More problematic is that a UKB basis often suffers from severe linear dependence. Moreover, UKB does not offer a faster convergence to the basis set limit than RKB.
\end{enumerate}
In short, RKB is the right choice for discretizing the Dirac equation in the absence of external magnetic fields.
Since RKB is also a built-in condition for two-component relativistic theories\cite{LiuMP}, it should be regarded as a cornerstone of relativistic quantum chemistry.
In the presence of external magnetic fields, RKB can be generalized to
\begin{eqnarray}
Z=\begin{pmatrix}Z_{11}&Z_{12}\\ Z_{21}&Z_{22}\end{pmatrix},\label{Zop}
\end{eqnarray}
which leads to a most general expansion of $\psi_p$,
\begin{eqnarray}
  \psi_p &=& Z\tilde{\psi}_p, \quad
  \tilde{\psi}_p =
  \begin{pmatrix}
    g_\mu A_{\mu p} \\
    g_\mu B_{\mu p}
  \end{pmatrix}\nonumber \\
  &=&\begin{pmatrix}
    Z_{11}g_\mu A_{\mu p}+Z_{12}g_\mu B_{\mu p} \\
    Z_{21}g_\mu A_{\mu p}+Z_{22}g_\mu B_{\mu p}
  \end{pmatrix}.\label{eq:psi}
\end{eqnarray}
Specific examples for the $Z$ operator \eqref{Zop} can be found from Refs. \citenum{PhysRep} and \citenum{TCANMR} and are not repeated here.
\section{The $\mbox{eQED}$ Hamiltonian}\label{SeceQED}
Given the one-electron Dirac operator, the question is how to construct a relativistic many-electron Hamiltonian.
The common practice is to add in simply the Coulomb interaction.
Since the instantaneous Gaunt and Breit interactions can also be derived in a semiclassical manner\cite{Dyallbook},
they can likewise be included, thereby leading to the Dirac-Coulomb-Gaunt/Breit (DC/DCG/DCB) Hamiltonian
\begin{eqnarray}
H &=& \sum_{i=1}^N D(i) + \frac{1}{2}\sum_{i\ne j}^NV(r_{ij}),\label{HDCB}\\
D&=&c\boldsymbol{\alpha}\cdot\boldsymbol{p}+(\beta-1)mc^2-\sum_{A}^{N_A}\frac{Z_A}{|\boldsymbol{R}_A-\boldsymbol{r}|},\\
V(r_{12})&=&V_C(r_{12}) + V_B(r_{12}),\\
V_C(r_{12})&=&\frac{1}{r_{12}},\\
V_B(r_{12})&=&V_G(r_{12})+V_g(r_{12}),\label{BreitInt}\\
V_G(r_{12})&=&-\frac{\boldsymbol{\alpha}_i\cdot\boldsymbol{\alpha}_j}{r_{12}},\label{GauntInt}\\
V_g(r_{12})&=&\frac{\boldsymbol{\alpha}_1\cdot\boldsymbol{\alpha}_2}{2r_{12}}-
\frac{(\boldsymbol{\alpha}_1\cdot\boldsymbol{r}_{12})(\boldsymbol{\alpha}_2\cdot\boldsymbol{r}_{12})}{2r^3_{12}}.
\end{eqnarray}
On the formal side, the DCB Hamiltonian should be adopted as it is correct to $\mathcal{O}(\alpha^2)$, whereas so is neither DC nor DCG.
Yet, on the practical side, the DCG Hamiltonian is more appealing for it describes all inter-electronic spin-same-orbit, spin-other-orbit, orbit-orbit, and spin-spin interactions
of $\mathcal{O}(\alpha^2)$ and is computationally cheaper than DCB. That is,
the difference between DCB and DCG is merely a scalar gauge term $V_g$ that is of minor importance but leads to complicated integrals.
As such, the DC and DCG Hamiltonians have been the major basis of relativistic quantum chemistry
for molecular chemistry and physics. However, unlike the Schr{\"o}dinger-Coulomb (SC) Hamiltonian that has well-defined mathematical and spectral properties, such \emph{ad hoc}
relativistic Hamiltonians have serious problems\cite{PCCPNES,eQED}. Without going into details, suffice it to say here
that such \emph{first-quantized} Hamiltonians violate a fundamental law of relativistic quantum mechanics,
viz., it is the charge instead of the number of particles that is conserved. Therefore, it is pointless to
solve the DC/DCG/DCB equation $H\Psi=E\Psi$ exactly,
unless one is interested in its mathematical solutions. Instead, to conserve the number of electrons, it is only consistent to adopt
the no-pair approximation (NPA) from the outset, regardless of the existence of bound states or not.
An immediate consequence is that the resulting energy $E_{\mathrm{np}}$ is not unique but is always dependent on how the projection operator is defined.
Since the projector can only be defined in terms of the PES of some effective potential, it can be said that $E_{\mathrm{np}}$
is always potential dependent, a situation that is very different from the FCI (full configuration interaction) solution of the Schr{\"o}dinger equation.
Even though such ambiguity can largely be removed by optimizing the potential/projector at a correlated level (e.g., no-pair full multiconfiguration self-consistent field
including orbital rotations to the unoccupied NES\cite{CorrelatedProjector}),
how to account for the (dynamic) correlation of NES still remains to be resolved. This requires a ``with-pair relativistic Hamiltonian'' in the first place.

\begin{figure}
\centering
\begin{center}
\includegraphics[width=0.7\columnwidth,keepaspectratio=true]{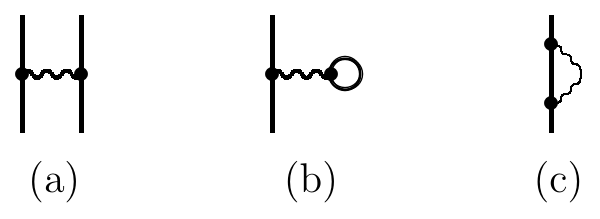}
\caption{Diagrammatical representation of the one-photon exchange, vacuum polarization and electron self-energy}
\label{FigureVPESE}
\end{center}
\end{figure}

As emphasized in the Introduction, the correct description of relativistic electrons must be done via second quantization.
More specifically, it is the ``extended hole theory'', the field Dirac picture coupled with
charge conjugation, that is the proper tool for constructing many-electron relativistic Hamiltonians. To begin with,
a \emph{primitive} second quantization of the Dirac matter field can be introduced, viz.,
\begin{eqnarray}
\hat{\phi}(\boldsymbol{r})=a_p\psi_p(\boldsymbol{r}),\quad a_p|vac\rangle=0,\quad p \in \mbox{PES, NES},
\end{eqnarray}
where the spinors are eigenfunctions of the following effective Dirac equation
\begin{eqnarray}
(D+U)\psi_p=\epsilon_p\psi_p,\label{UDEQ}
\end{eqnarray}
with $U$ being some local or nonlocal screening potential.
The term `primitive' here means that this form of second quantization does not distinguish the empty from the filled Dirac picture.
This gives rise to the following normal-ordered, second-quantized DC/DCG/DCB Hamiltonian
\begin{eqnarray}
\mathcal{H} &=&D_p^qa^p_q +\frac{1}{2}g_{pq}^{rs}a^{pq}_{rs},\quad p, q, r, s \in \mbox{PES, NES},\label{Hbase}\\
D_p^q&=&\langle\psi_p|D|\psi_q\rangle,\quad g_{pq}^{rs}=\langle\psi_p\psi_q|V(r_{12})|\psi_r\psi_s\rangle,\label{Gpqrs}\\
a^p_q&=&a^pa_q,\quad a^{pq}_{rs}=a^pa^qa_sa_r.
\end{eqnarray}
The filled Dirac picture can be realized in a finite basis representation by
setting the Fermi level below the energetically lowest of the $\tilde{N}$ ($=N_{e^-}$) occupied NES. The physical energy of an $N$-electron state can be calculated\cite{PCCPNES} as
the difference between those of states $\Psi(N;\tilde{N})$ and $\Psi(0;\tilde{N})$,
\begin{eqnarray}
E&=&\langle \Psi(N;\tilde{N})|\mathcal{H}|\Psi(N;\tilde{N})\rangle - \langle \Psi(0;\tilde{N})|\mathcal{H}|\Psi(0;\tilde{N})\rangle,\label{EFStot}
\end{eqnarray}
provided that the charge-conjugation symmetry is incorporated properly.
To do so, we first shift the Fermi level just above the top of the NES. This amounts to normal ordering
the Hamiltonian $\mathcal{H}$ \eqref{Hbase} with respect to the non-interacting vacuum $|0;\tilde{N}\rangle$ ($=|0_{e^-}; N_{e^-}\rangle$) of zero
positive energy electrons and $\tilde{N}$ negative-energy electrons.
Here, the CCC \eqref{CCC} of fermion operators\cite{eQED}  must be invoked, so as to obtain
\begin{eqnarray}
a^p_q&=&\{a^pa_q\}_n+\langle 0;\tilde{N}|\frac{1}{2}[a^p,a_q]|0;\tilde{N}\rangle,\quad p, q \in \mbox{PES, NES},\\
&=&\{a^pa_q\}_n-\frac{1}{2}\delta^p_q \sgn(\epsilon_q), \quad p, q \in \mbox{PES, NES}, \label{OneCCC1}\\
D_p^qa^p_q&=&D_p^q\{a^pa_q\}_n + C_{1n}, \quad C_{1n} = -\frac{1}{2}D_p^p\sgn(\epsilon_p),\label{OneCCC2}
\end{eqnarray}
where the subscript $n$ of the curly brackets emphasizes that the normal ordering is taken with respect to the reference $|0;\tilde{N}\rangle$.
More specifically,
\begin{eqnarray}\label{Normal-aop}
\{a^pa_q\}_n&=&
\begin{cases}
a^pa_q,  \quad \epsilon_p>0, \epsilon_q>0,\\
a^pa_q,  \quad \epsilon_p>0, \epsilon_q<0,\\
a^pa_q,  \quad \epsilon_p<0, \epsilon_q>0,\\
-a_qa^p, \quad \epsilon_p<0, \epsilon_q<0.
\end{cases}
\end{eqnarray}
By applying the relation \eqref{CCC} repeatedly we obtain
\begin{eqnarray}
a^{pq}_{rs}&=&\{a^{pq}_{rs}\}_n-\frac{1}{2}\{ \delta^p_ra^q_s \sgn(\epsilon_r)+ \delta^q_sa^p_r \sgn(\epsilon_s)
-\delta^q_ra^p_s\sgn(\epsilon_r)-\delta^p_sa^q_r\sgn(\epsilon_s)\}_n\nonumber\\
& &+\frac{1}{4}(\delta^p_r\delta^q_s-\delta^q_r\delta^p_s) \sgn(\epsilon_r)\sgn(\epsilon_s),\label{2CCC}
\end{eqnarray}
and hence
\begin{eqnarray}
\frac{1}{2}g_{pq}^{rs}a^{pq}_{rs}&=&\frac{1}{2}g_{pq}^{rs}\{a^{pq}_{rs}\}_n+Q_p^q\{a^p_q\}_n +C_{2n}, \label{TwoCCC1}\\
Q_p^q&=&\tilde{Q}_p^q+\bar{Q}_p^q=-\frac{1}{2}\bar{g}_{ps}^{qs}\sgn(\epsilon_s),\label{QDef}\\
\tilde{Q}_p^q&=&-\frac{1}{2}g_{ps}^{qs}\sgn(\epsilon_s),\label{QVP}\\
\bar{Q}_p^q &=& \frac{1}{2}g_{ps}^{s q}\sgn(\epsilon_s),\label{QSE}\\
C_{2n}&=&\frac{1}{8}\bar{g}_{pq}^{pq}\sgn(\epsilon_p)\sgn(\epsilon_q)=-\frac{1}{4}Q_p^p\sgn(\epsilon_p).\label{TwoCCC2}
\end{eqnarray}
Note that the implicit summations in $C_{1n}$ \eqref{OneCCC2}, ${\tilde Q}$ \eqref{QVP}, ${\bar Q}$ \eqref{QSE}, and $C_{2n}$ \eqref{TwoCCC2} include all the PES and NES, whether occupied or not.
The Hamiltonian $\mathcal{H}$ \eqref{Hbase} in the filled Dirac picture can then be written as
\begin{eqnarray}
\mathcal{H}&=&H_a^{\mathrm{QED}}+ C_n,\label{H2}\\
H_a^{\mathrm{QED}}&=&H^{\mathrm{FS}}_a + Q_p^q\{a^p_q\}_n\label{HnH},\\
H^{\mathrm{FS}}_a&=&D_p^q\{a^p_q\}_n +\frac{1}{2}g_{pq}^{rs}\{a^{pq}_{rs}\}_n,\label{KZFS}\\
C_n&=& C_{1n}+C_{2n} =\langle 0;\tilde{N}|H|0;\tilde{N}\rangle
   =-\frac{1}{2}D_p^p\sgn(\epsilon_p)-\frac{1}{4}Q_p^p\sgn(\epsilon_p).\label{HnC}
\end{eqnarray}
$H_a^{\mathrm{QED}}$ \eqref{HnH} is just the desired ``with-pair relativistic Hamiltonian'' or simply effective QED (eQED) Hamiltonian, whereas
$H^{\mathrm{FS}}_a$ is the so-called Fock space Hamiltonian advocated by Kutzelnigg\cite{KutzFS} (see also Ref. \citenum{Dyallbook}),
which missed, by construction, the vacuum polarization (VP) $\tilde{Q}$ \eqref{QVP} (see Fig. \ref{FigureVPESE}(b))  and electron self-energy (ESE) $\bar{Q}$ \eqref{QSE} (see Fig. \ref{FigureVPESE}(c)).
If wanted, $H_a^{\mathrm{QED}}$ \eqref{HnH} can also be expressed in terms of the $b$-operators in view of the relations \eqref{a2b-op}, viz.,
\begin{eqnarray}
H_b^{\mathrm{QED}}&=&(D+Q)_p^q \{b^p b_q\} + (D+Q)_{\tilde{p}}^q \{b_{\tilde{p}}b_q\}+(D+Q)_p^{\tilde{q}}\{b^pb^{\tilde{q}}\}+(D+Q)_{\tilde{p}}^{\tilde{q}}\{b_{\tilde{p}}b^{\tilde{q}}\}\nonumber\\
&+&\frac{1}{4}\bar{g}_{pq}^{rs}\{b^pb^qb_sb_r\}+\frac{1}{2}\bar{g}_{\tilde{p}q}^{rs}\{b_{\tilde{p}}b^qb_sb_r\}+\frac{1}{2}\bar{g}_{pq}^{\tilde{r}s}\{b^pb^qb_sb^{\tilde{r}}\}
+\frac{1}{4}\bar{g}_{\tilde{p}\tilde{q}}^{rs}\{b_{\tilde{p}}b_{\tilde{q}}b_sb_r\}\nonumber\\
&+&\frac{1}{4}\bar{g}_{pq}^{\tilde{r}\tilde{s}}\{b^pb^qb^{\tilde{s}}b^{\tilde{r}}\}+\bar{g}_{\tilde{p}q}^{\tilde{r}s}\{b_{\tilde{p}}b^qb_sb^{\tilde{r}}\}
+\frac{1}{2}g_{\tilde{p}\tilde{q}}^{\tilde{r}s}\{b_{\tilde{p}}b_{\tilde{q}}b_sb^{\tilde{r}}\}\nonumber\\
&+&\frac{1}{2}\bar{g}_{p\tilde{q}}^{\tilde{r}\tilde{s}}\{b^pb_{\tilde{q}}b^{\tilde{s}}b^{\tilde{r}}\} +\frac{1}{4}\bar{g}_{\tilde{p}\tilde{q}}^{\tilde{r}\tilde{s}}\{b_{\tilde{p}}b_{\tilde{q}}b^{\tilde{s}}b^{\tilde{r}}\},\label{HbH}
\end{eqnarray}
where the normal ordering is taken with respect to $|0\rangle$. Note that the eQED Hamiltonian \eqref{HnH}/\eqref{HbH}
can also be obtained by a diagrammatical procedure\cite{IJQCeQED,eQEDBook2017}.
Had the standard contraction \eqref{NRCCC2} of fermion operators
been taken, we would obtain the following ``configuration space'' (CS) Hamiltonian
\begin{eqnarray}
H_a^{\mathrm{CS}}=H_a^{\mathrm{FS}}+ \bar{g}_{p\tilde{j}}^{q\tilde{j}}\{a^p_q\}_n. \label{HnCS}
\end{eqnarray}
At variance with the $Q$ potential \eqref{QDef},
the potential $\bar{g}_{p\tilde{j}}^{q\tilde{j}}$ here arises from the occupied NES $\{\psi_{\tilde j}\}$ alone (which is also
a conventional interpretation of Fig. \ref{FigureVPESE}(b)). It is infinitely repulsive,
leading to that no atom would be stable. This is of course plainly wrong.

Finally, the proper evaluation of the $Q$ potential \eqref{QDef} should be discussed.
The Coulomb-only $\tilde{Q}$ term \eqref{QVP} is the full vacuum polarization\cite{LindgrenWKterm1993} due to
the polarization density $\rho_{vp}$ \eqref{SDenvp}. In practice, it can be split into the
Uehling \cite{UehlingVP} and Wichmann-Kroll \cite{WichmannVP} terms, which can then be evaluated
with the analytic formulae\cite{UehlingVPformular,WichmannVPformula}.
The ESE term $\bar{Q}$ \eqref{QSE} is more difficult to handle.
In addition to the Coulomb interaction, the transverse-photon contribution should also be included.
In the Coulomb gauge adopted here, the transverse part of the ESE reads\cite{CEOBook}
\begin{eqnarray}
(\bar{Q}_T)_p^q&=&\frac{1}{2}\langle p| \Sigma_T^C(\epsilon_p) + \Sigma_T^C(\epsilon_q)|q\rangle,\\
\langle p|\Sigma_T^C(\epsilon_p)|q\rangle&=&\langle ps|
\int_0^{\infty} \frac{cdk f_T^C(k, r_1, r_2)}{\epsilon_p-\epsilon_{s}-(ck-i\gamma)\sgn(\epsilon_{s})}
|s q\rangle,\\
f_T^C(k, r_1, r_2)&=&\frac{{\rm sin}(k r_{12})}{\pi r_{12}}[\boldsymbol{\alpha}_1\cdot\boldsymbol{\alpha}_2
-\frac{(\boldsymbol{\alpha}_1\cdot\boldsymbol{\nabla}_1)(\boldsymbol{\alpha}_2\cdot\boldsymbol{\nabla}_2)}{k^2}].
\end{eqnarray}
Therefore, the total ESE (still denoted as $\bar{Q}$) can be written in a symmetric form
\begin{eqnarray}
\bar{Q}_p^q&=&\frac{1}{2}\langle p|\Sigma_C^C + \Sigma_T^C(\epsilon_p) + \Sigma_T^C(\epsilon_q)|q\rangle,\quad \langle p|\Sigma_C^C|q\rangle=g_{ps}^{sq}\sgn(\epsilon_{s}).\label{QSETot}
\end{eqnarray}
It has recently been shown that the ESE \eqref{QSETot} can be fitted into a simple and accurate semilocal model operator for each atom\cite{ShabaevModelSE,ShabaevModelSEcode2018}.
Therefore, the VP-ESE (Lamb shift) can readily be included in the mean-field treatment, so as to account for
screening effects on the VP-ESE automatically.

\section{Application of the $\mbox{eQED}$ Hamiltonian}\label{SecNES}
In this section, the occupied PES and NES are to be denoted respectively by $\{i, j, \cdots \}$ and $\{\tilde{i}, \tilde{j}, \cdots\}$,
whereas the unoccupied PES by $\{a, b, \cdots\}$. Unspecified orbitals are denoted as $\{p, q, r, s\}$.
When necessary, the NES will explicitly be designated by $\{\tilde{p}, \tilde{q}, \tilde{r}, \tilde{s}\}$.
\subsection{The second-order QED energy of an $N$-electron system}\label{SecE2}
The eQED Hamiltonian $H_a^{\mathrm{QED}}$ \eqref{HnH} or $H_b^{\mathrm{QED}}$ \eqref{HbH} can be employed in the Bloch equation to determine
the wave operators order by order. The resulting energy expressions are in full agreement with those obtained by the S-matrix
formulation of QED\cite{eQED}. However, the procedure treating all the PES as particles is rather involved.
It is more expedite\cite{PCCPNES} to calculate the physical energy
according to Eq. \eqref{EFStot} by treating the occupied PES also as holes. That is, to calculate the first term on the right-hand side of  Eq. \eqref{EFStot},
the eQED Hamiltonian $H_a^{\mathrm{QED}}$ \eqref{HnH}
can further be normal-ordered with respect to the non-interacting reference $|N;\tilde{N}\rangle$, the zero order of $\Psi(N;\tilde{N})$.
Since the normal ordering is now taken with respect to the occupied PES alone,
the standard contraction of Fermion operators, e.g.,
\begin{eqnarray}
\bcontraction[0.5ex]{}{a^p}{}{a_q}a^pa_q=\langle N;\tilde{N}|\{a^pa_q\}_n|N;\tilde{N}\rangle=\langle N;0|a^pa_q|N;0\rangle=\delta^p_q n_q, \quad \epsilon_q>0, \label{NRCCC}
\end{eqnarray}
should be invoked. More specifically,
\begin{eqnarray}
D_p^q\{a^p_q\}_n&=& D_p^q\{a^p_q\}_F + D_i^i,\\
Q_p^q\{a^p_q\}_n&=& Q_p^q\{a^p_q\}_F + Q_i^i,\\
\frac{1}{2}g_{pq}^{rs}\{a^{pq}_{rs}\}_n&=&\frac{1}{2}g_{pq}^{rs}\{a^{pq}_{rs}\}_F + (V_{HF})_p^q\{a^p_q\}_F+\frac{1}{2}(V_{HF})_i^i.
\end{eqnarray}
Consequently, we have\cite{PhysRep}
\begin{eqnarray}
\mathcal{H}&=&H_F^{\mathrm{QED}}+ C_F,\label{H3}\\
H_F^{\mathrm{QED}}&=&f_p^q\{a^p_q\}_F + \frac{1}{2}g_{pq}^{rs}\{a^{pq}_{rs}\}_F,\label{HeQED}\\
f_p^q&=&(f_e)_p^q+Q_p^q,\label{full-Fock}\\
(f_e)_p^q&=&D_p^q+(V_{HF})_p^q,\quad (V_{HF})_p^q=\bar{g}_{pj}^{qj},\label{Feop}\\
C_F&=&C_n+E^{[1]}=(D+V_{HF}+Q)_i^i-(\frac{1}{2}D+\frac{1}{4}Q)_p^p\sgn(\epsilon_p),\\
E^{[1]}&=&E^{[1]}_{\mathrm{np}}+Q_i^i,\label{E1full}\\
E^{[1]}_{\mathrm{np}}&=&(D+\frac{1}{2}V_{HF})_i^i.\label{E1np}
\end{eqnarray}
To facilitate the use of many-body perturbation theory (MBPT) for electron correlation, the Hamiltonian \eqref{H3}
can further be partitioned as
\begin{eqnarray}
\mathcal{H}&=& H_{0A}+V_{0A}+V_{1A}+V_{2A},\label{HA}\\
H_{0A}&=&\epsilon_p\{a^p_p\}_F+\sum_i^N\epsilon_i-\frac{1}{2}\epsilon_p\sgn(\epsilon_p),\label{H0A}\\
V_{0A}&=&(Q-U+\frac{1}{2}V_{HF})_i^i+(\frac{1}{2}U-\frac{1}{4}Q)_p^p\sgn(\epsilon_p),\label{V0A}\\
V_{1A}&=&(V_{1A})_p^q\{a^p_q\}_F, \quad (V_{1A})_p^q=(Q+V_{HF}-U)_p^q,\label{V1A}\\
V_{2A}&=& \frac{1}{2}g_{pq}^{rs}\{a^{pq}_{rs}\}_F,\label{V2A}
\end{eqnarray}
where the appearance of the counter potential
\begin{eqnarray}
-U_p^q\{a^p_q\}&=&-U_p^q\{a^p_q\}_n+\frac{1}{2}U_p^p\sgn(\epsilon_p)\nonumber\\
&=&-U_p^q\{a^p_q\}_F-U_i^i+\frac{1}{2}U_p^p\sgn(\epsilon_p)
\end{eqnarray}
is due to the fact that the general mean-field equation \eqref{UDEQ} has been employed to determine
the spinors and energy levels.
%
As for the second term of Eq. \eqref{EFStot}, the Hamiltonian \eqref{H2} can be partitioned as
\begin{eqnarray}
\mathcal{H}&=& H_{0B}+V_{0B}+V_{1B}+V_{2B},\label{HB}\\
H_{0B}&=&\epsilon_p\{a^p_p\}_n-\frac{1}{2}\epsilon_p\sgn(\epsilon_p),\label{H0B}\\
V_{0B}&=&(\frac{1}{2}U-\frac{1}{4}Q)_p^p\sgn(\epsilon_p),\label{V0B}\\
V_{1B}&=&(V_{1B})_p^q\{a^p_q\}_n,\quad V_{1B}=Q-U, \label{V1B}\\
V_{2B}&=& \frac{1}{2}g_{pq}^{rs}\{a^{pq}_{rs}\}_n.\label{V2B}
\end{eqnarray}
Following the standard MBPT, we obtain immediately
\begin{align}
E^{(0)}  &=-\frac{1}{2}\epsilon_p\sgn(\epsilon_p-\epsilon_F)+\frac{1}{2}\epsilon_p\sgn(\epsilon_p)=\sum_i^N\epsilon_i,\label{Eh0}\\
E^{(1)}  &=V_{0A}-V_{0B}=(\frac{1}{2} V_{HF} -U + Q)_i^i,\label{Eh1}\\
E^{(2)}  &=E^{(2)}_1+E^{(2)}_2,\label{QEDE2}\\
E^{(2)}_1&=\left[\frac{(V_{1A})_i^a (V_{1A})_a^i}{\epsilon_i-\epsilon_a}+\frac{(V_{1A})_{\tilde i}^a (V_{1A})_a^{\tilde i}}{\epsilon_{\tilde i}-\epsilon_a}\right]\nonumber\\
         &-\left[\frac{(V_{1B})_{\tilde i}^i (V_{1B})_i^{\tilde i}}{\epsilon_{\tilde i}-\epsilon_i}+\frac{(V_{1B})_{\tilde i}^a (V_{1B})_a^{\tilde i}}{\epsilon_{\tilde i}-\epsilon_a}\right]\label{E21}\\
         &=E^{(2)}_{\rm{FS},1}+E^{(2)}_{\rm{Q},1},\\
E^{(2)}_{\rm{FS},1}&=\frac{(V_{HF}-U)_i^a(V_{HF}-U)_a^i}{\epsilon_i-\epsilon_a}+\frac{(V_{HF}-U)_{\tilde i}^a(V_{HF}-U)_a^{\tilde i}}{\epsilon_{\tilde i}-\epsilon_a}\nonumber\\
                   & -\frac{U_{\tilde i}^aU_a^{\tilde i}}{\epsilon_{\tilde i}-\epsilon_a}-\frac{U_{\tilde i}^iU_i^{\tilde i}}{\epsilon_{\tilde i}-\epsilon_i},\label{EFS1}\\
E^{(2)}_{\rm{Q},1}&=\frac{(V_{HF}-U)_i^aQ_a^i+Q_i^a(V_{HF}-U)_a^i+Q_i^aQ_a^i}{\epsilon_i-\epsilon_a}\nonumber\\
                  &+\frac{(V_{HF})_{\tilde i}^aQ_a^{\tilde i}+Q_{\tilde i}^a(V_{HF})_a^{\tilde i}}{\epsilon_{\tilde i}-\epsilon_a}
                   -\frac{Q_{\tilde i}^iQ_i^{\tilde i}-U_{\tilde i}^iQ_i^{\tilde i}-Q_{\tilde i}^iU_i^{\tilde i}}{\epsilon_{\tilde i}-\epsilon_i},\label{EVS1}
\end{align}
\begin{align}
E^{(2)}_2&=\frac{1}{4}\frac{\bar{g}_{mn}^{ab}\bar{g}_{ab}^{mn}}
                           {\epsilon_m+\epsilon_n-\epsilon_a-\epsilon_b}                  |_{m, n = i, j,\tilde{i},\tilde{j}}
          -\frac{1}{4}\frac{\bar{g}_{\tilde{i}\tilde{j}}^{pq}\bar{g}_{pq}^{\tilde{i}\tilde{j}}}
                           {\epsilon_{\tilde i}+\epsilon_{\tilde j}-\epsilon_p-\epsilon_q}|_{p, q=i, j, a, b}\label{E22Goldstone}\\
         &=\left[\frac{1}{4}\frac{\bar{g}_{ij}^{ab}\bar{g}_{ab}^{ij}} {\epsilon_i+\epsilon_j-\epsilon_a-\epsilon_b}
          +\frac{1}{2}\frac{\bar{g}_{i{\tilde j}}^{ab}\bar{g}_{ab}^{i{\tilde j}}} {\epsilon_i+\epsilon_{\tilde j}-\epsilon_a-\epsilon_b}\right]\nonumber\\
         &-\left[\frac{1}{4}\frac{\bar{g}_{{\tilde i}{\tilde j}}^{ij}\bar{g}_{ij}^{{\tilde i}{\tilde j}}} {\epsilon_{\tilde i}+\epsilon_{\tilde j}-\epsilon_i-\epsilon_j}
          +\frac{1}{2}\frac{\bar{g}_{{\tilde i}{\tilde j}}^{ia}\bar{g}_{ia}^{{\tilde i}{\tilde j}}} {\epsilon_{\tilde i}+\epsilon_{\tilde j}-\epsilon_i-\epsilon_a}\right].\label{E22}
\end{align}
\begin{figure}[t]
\centering
\begin{center}
\includegraphics[width=0.7\columnwidth,keepaspectratio=true]{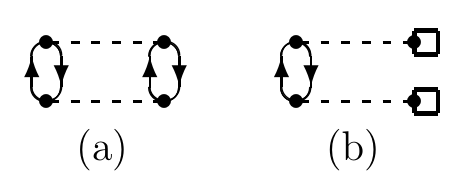}
\caption{Diagrammatical representation of the second-order QED energy. (a)
antisymmetrized two-body; (b) one-body. For the $\Psi(N;\tilde{N})$
state, the particles (up-going lines) and holes (down-going lines) are $\{a, b\}$ and
$\{i, j, \tilde{i}, \tilde{j}\}$, respectively, and the one-body potential represented by the square is
$V_{1A}$. For the $\Psi(0;\tilde{N})$ state, the particles and holes are $\{a, b, i, j\}$ and
$\{\tilde{i}, \tilde{j}\}$, respectively, and the one-body potential is $V_{1B}$. A global negative sign
should be inserted to the terms of $\Psi(0;\tilde{N})$. 
}\label{GoldstoneE2}
\end{center}
\end{figure}
The first and second terms of $E^{(2)}_1$ \eqref{E21} and $E^{(2)}_2$ \eqref{E22Goldstone} arise from the $\Psi(N;\tilde{N})$ and $\Psi(0;\tilde{N})$ states, respectively.
The one-body $E^{(2)}_1$ \eqref{E21} can further be decomposed into two terms, $E^{(2)}_{\mathrm{FS},1}$ \eqref{EFS1} and $E^{(2)}_{\mathrm{Q},1}$ \eqref{EVS1}.
Both $E^{(2)}_{\mathrm{FS},1}$ \eqref{EFS1} and $E^{(2)}_2$ \eqref{E22} arise from the Fock space Hamiltonian\cite{KutzFS} $H^{\mathrm{FS}}_a$ \eqref{KZFS},
while $E^{(2)}_{\mathrm{Q},1}$ \eqref{EVS1} is due exclusively to the VP and ESE [NB: $E^{(2)}_{\mathrm{Q},1}$ and
the $Q$ term in $E^{(1)}$ will not show up if the $Q$ potential is included in the mean-field equation \eqref{effDEQ}].
The two terms of $E^{(2)}$ \eqref{QEDE2} can be represented by the same Goldstone-like diagrams shown in Fig. \ref{GoldstoneE2}.
It is just that the particles and holes, as well as the one-body potential, are interpreted differently.
Note that
the frequency-dependent Breit interaction
\begin{eqnarray}
V_T(\omega,r_{12})&=&-\boldsymbol{\alpha}_1\cdot\boldsymbol{\alpha}_2 \frac{\cos(|q|r_{12})}{r_{12}},\quad \omega=qc \nonumber\\
&+&[(\boldsymbol{\alpha}_1\cdot\boldsymbol{\nabla}_1),[(\boldsymbol{\alpha}_2\cdot\boldsymbol{\nabla}_2),
\frac{\cos(|q|r_{12})-1}{q^2r_{12}}]] \label{VTZ}
\end{eqnarray}
must be employed to account for the contribution of NES to correlation,
which would otherwise be severely overestimated if the frequency dependence is neglected\cite{SCF-Breit}. This amounts to replacing the integrals $g_{pq}^{rs}$ in $E^{(2)}_2$ \eqref{E22} with
\begin{eqnarray}
g_{pq}^{rs}&=&\langle p q |V_C(r_{12}) + \frac{1}{2} V_T(\epsilon_r-\epsilon_p,r_{12})+\frac{1}{2}V_T(\epsilon_s-\epsilon_q,r_{12})|rs\rangle.\label{Gtrans}
\end{eqnarray}
\begin{figure}[t]
\centering
\begin{center}
\includegraphics[width=0.6\columnwidth,keepaspectratio=true]{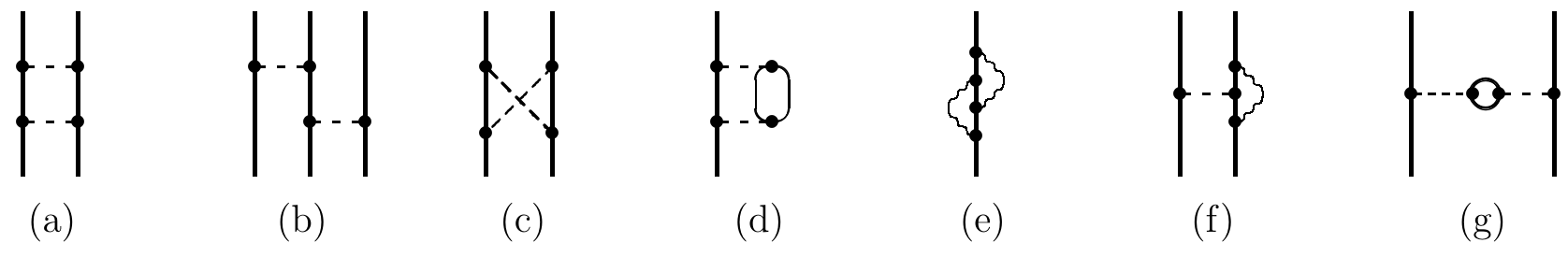}
\caption{Feynman diagrams for $E^{(2)}_2$.
}\label{FigureE2Feynman}
\end{center}
\end{figure}
Here we should recall again the most recent and to date most accurate relativistic calculations\cite{SchwerdtfegerAuPRL2017} of the IP and EA of Au:
the calculated IP (9.2288 eV) deviates from the experimental one (9.2256 eV) somewhat larger than the corresponding EA (calculated 2.3072 vs. experimental 2.3086 eV). This is counterintuitive, since IPs can usually be calculated more accurately than EAs. There could be two possible reasons for this:
(a) the no-pair correlation still need to be improved and (b) the missing contribution of NES to correlation has to be taken into account. Even if the contribution of NES to correlation is not the reason for such discrepancy, it is certainly important for core properties.

It also deserves to be mentioned that the same $E^{(2)}_2$ \eqref{E22Goldstone} would correspond
to the seven Feynman diagrams shown in Fig. \ref{FigureE2Feynman}, if the
occupied PES are to be treated as particles instead of holes as in diagram Fig. \ref{GoldstoneE2}(a). The first three and the next four
of these Feynman diagrams are usually called non-radiative and radiative contributions in QED but all of which are contributions to electron correlation in the present context. It is now clear that only the $Q$ potential \eqref{QDef} entering the eQED Hamiltonian \eqref{HnH}/\eqref{HbH} arises from the unconentional contraction \eqref{CCC} of fermion operators, whereas the treatment of electron correlation follows standard many-body theories in conjunction with the filled Dirac picture. This is because the $n$-th order correlation energy $E^{(n)}=\langle vac|V\Omega^{(n-1)}|vac\rangle$ arises from the (full) contraction between the first-order fluctuation potential $V$ and the $(n-1)$th-order wave operator $\Omega^{(n-1)}$ that are already normal ordered separately and hence originate from ``different times''. In the parlance of diagrams, all the terms herein refer to reducible Feynman diagrams (see Fig. \ref{FigureE2Feynman}). In other words, only those irreducible
Feynman diagrams that go beyond the eQED Hamiltonian (which is, by definition, linear in the two-particle interaction involving only one-photon
exchange) must be treated via full QED, a time-dependent perturbation theory.

Finally, it is instructive to compared the second-order QED energy $E^{(2)}$ \eqref{QEDE2} with that of the configuration space approach\cite{PCCPNES}
\begin{align}
E^{(2)}_{\rm CS}&=E^{(2)}_{{\rm CS},1}+E^{(2)}_{{\rm CS},2},\label{E2CS}\\
E^{(2)}_{{\rm CS},1}&=\frac{(V_{HF}-U)_i^a(V_{HF}-U)_a^i}{\epsilon_i-\epsilon_a}
+\frac{(V_{HF}-U)_i^{\tilde i}(V_{HF}-U)_{\tilde i}^i}{\epsilon_i-\epsilon_{\tilde i}},\label{E2CS1}\\
E^{(2)}_{{\rm CS},2}&=\frac{1}{4}\frac{\bar{g}_{ij}^{ab}\bar{g}_{ab}^{ij}}
{ \epsilon_i+\epsilon_j-\epsilon_a-\epsilon_b }
+\frac{1}{4}\frac{\bar{g}_{ij}^{{\tilde i}{\tilde j}}\bar{g}_{{\tilde i}{\tilde j}}^{ij}}
{ \epsilon_i+\epsilon_j-\epsilon_{\tilde i}-\epsilon_{\tilde j} }
+\frac{1}{2}\frac{\bar{g}_{ij}^{ a{\tilde j}}\bar{g}_{a{\tilde j} }^{ij}}
{ \epsilon_i+\epsilon_j-\epsilon_a -\epsilon_{\tilde j}}.\label{E2CS2}
\end{align}
It is seen that $E^{(2)}_{{\rm CS},1}$ and $E^{(2)}_{{\rm CS},2}$ agree respectively with $E^{(2)}_{{\rm FS},1}$ \eqref{EFS1} and
$E^{(2)}_2$ \eqref{E22} only in the first terms involving solely the PES, but are very different from the latter in the terms involving the NES.
In particular, the denominator of the last term of Eq. \eqref{E2CS2} can be zero (e.g., $\epsilon_a=\epsilon_i+|X|$ and $\epsilon_{\tilde j}=\epsilon_j-|X|$). Since there exists an infinite number of such ``$+-$'' intermediates, this problem has been termed continuum dissolution\cite{Brown-Ravenhall}. As shown here, it is purely an artefact due to the underlying empty Dirac picture.
As such, only no-pair projected wave functions are acceptable in the configuration space formulation. Efforts\cite{Nakatsuji-DC} to solve exactly the DC/DCG/DCB equation
$H\Psi=E\Psi$ are then purely mathematical exercises. Since there is no analytic Hamiltonian in Fock space as well, the term ``exact (analytic) relativistic wave function'' is simply meaningless\cite{RelR12}.


\subsection{Mean-field theory of real positrons}\label{SecPositron}
As an another application of the eQED Hamiltonian \eqref{HnH}, we present here a mean-field theory for
a system of $N$ electrons and $\tilde{M}$ positrons. The energy up to first order reads
\begin{eqnarray}
E_{\mathrm{ep}}^{[1]}
&=&\langle N;\tilde{N}|A^\dag H_a^{\mathrm{QED}} A|N;\tilde{N}\rangle, \quad A = \Pi_{\tilde i}^{\tilde M} a_{\tilde i}\label{E1-2}\\
&=& [\sum_{i=1}^N(D+Q)_i^i + \frac{1}{2} \sum_{i,j=1}^N\bar{g}_{ij}^{ij}] + [-\sum_{\tilde{i}=\tilde{1}}^{\tilde M} (D+Q)_{\tilde i}^{\tilde i}
+ \frac{1}{2} \sum_{\tilde{i},\tilde{j}=\tilde{1}}^{\tilde M} \bar{g}_{\tilde{i}\tilde{j}}^{\tilde{i}\tilde{j}}]\nonumber\\
&&-\sum_{i=1}^N\sum_{\tilde{j}=\tilde{1}}^{\tilde M} \bar{g}_{i\tilde{j}}^{i\tilde{j}}.\label{E1-3}
\end{eqnarray}
Use of Wick's theorem for expressing products of normal-ordered operators as a linear combination of contracted ones
has been made when going from Eq. \eqref{E1-2} to \eqref{E1-3}. The first and second terms are the average
energies of the $N$ electrons and of the $\tilde{M}$ positrons, respectively, whereas the third, cross term represents their mutual interaction.
The negative sign in the second and third terms results from the normal ordering \eqref{Normal-aop} implicit in $H_a^{\mathrm{QED}}$ \eqref{HnH},
and can be understood as a negative occupation number ($n_{\tilde{i}}=-1$) of the hole arising from the ionization $a_{\tilde{i}}|N;\tilde{N}\rangle$.
The other occupied NES not involved in the ionization $\Pi_{\tilde i}^{\tilde M} a_{\tilde i}|N;\tilde{N}\rangle$ have been normal-ordered away,
and can therefore be viewed as unoccupied, just like the unoccupied PES.
As such, the expression \eqref{E1-3} can be written as
\begin{eqnarray}
E_{\mathrm{ep}}^{[1]}&=&\sum_k n_k (D+Q)_k^k + \frac{1}{2}\sum_{k,l} n_k n_l \bar{g}_{kl}^{kl}, \quad k,l \in \mathrm{PES, NES}, \label{E1ep}
\end{eqnarray}
by assigning an occupation number $n_k$ to each orbital $\psi_k$:
$n_k$ is zero for the unoccupied PES and NES, $+1$ for the $N$ occupied PES, and $-1$ for the $\tilde{M}$ occupied NES.
Formally, this agrees with the empty Dirac picture. However, such agreement between the empty and filled Dirac pictures
holds only at the mean-field level but not at the correlated level (see Sec. \ref{SecE2}). More generally,
such agreement holds for all one-body but not for any two-body operators\cite{PCCPNES}.

To minimize the energy $E_{\mathrm{ep}}^{[1]}$ \eqref{E1ep} subject to the orthonormal conditions, we can introduce the following canonical
Lagrangian
\begin{eqnarray}
L=E_{\mathrm{ep}}^{[1]} -\sum_{k}n_k [\langle \psi_k|\psi_k\rangle-1]\epsilon_{k}, \quad k \in \mathrm{PES, NES}.
\end{eqnarray}
The condition $\frac{\delta L}{\delta \psi_i^\dag}=0$ then gives rise to
\begin{eqnarray}
f n_i|\psi_i\rangle &=& \epsilon_{i}n_i|\psi_i\rangle,\quad i\in \mathrm{PES, NES},\label{fpositron2}
\end{eqnarray}
where
\begin{eqnarray}
f&=& D + Q + \sum_k n_k \bar{g}_{k\cdot}^{k\cdot},\quad k \in \mathrm{PES, NES},\label{fpositron}\\
f_p^q&=&(D + Q)_p^q + \sum_k n_k \bar{g}_{kp}^{kq}, \quad k \in \mathrm{PES, NES}.\label{fpositron1}
\end{eqnarray}
As it stands, Eq. \eqref{fpositron2} determines only the occupied PES and NES but which can be extended to the unoccupied ones (which are arbitrary anyway), viz.,
\begin{eqnarray}
f |\psi_p\rangle &=& \epsilon_{p}|\psi_p\rangle,\quad p\in \mathrm{PES, NES}.\label{fpositron1}
\end{eqnarray}
The energetically lowest PES and highest NES are to be occupied in each iteration.

Some remarks are in order: (a) the cross, exchange term $-\sum_{i\tilde{j}}g_{i\tilde{j}}^{\tilde{j}i}$ vanishes
in the nrl, meaning that electrons and positrons are distinguishable particles in the nonrelativistic world, such that their mutual anti-symmetrization is no longer required. In other words, only QED treats electrons and positrons on an equal footing. (b)
If the VP-ESE term $Q$ is neglected, the present mean-field theory of electrons and positrons will reduce to that formulated by Dyall in a different way\cite{DyallPositron}.
\section{No-pair relativistic Hamiltonians}\label{Sec2C}
There have been a number of comprehensive reviews\cite{LiuMP,SaueRev,ReiherRev,PhysRep,X2C2016,X2CBook2017} on
the no-pair relativistic Hamiltonians. Therefore, only a brief summary of the essentials is necessary here.
The no-pair relativistic Hamiltonians can be classified into four-component (4C), quasi-four-component (Q4C) and two-component (2C) ones,
the last of which can further be classified into approximate (A2C) and exact (X2C) two-component ones.
\subsection{Four-component}
First of all, confining the orbital indices of $H_a^{\mathrm{QED}}$ \eqref{HnH} only to PES leads to the following no-pair QED Hamiltonian
\begin{eqnarray}
H_{+}^{\rm{QED}}&=&D_p^qa^p_q+Q_p^q a^p_q+ \frac{1}{2}g_{pq}^{rs}a^{pq}_{rs}, \quad p, q, r, s \in \mbox{ PES} \label{npQED}\\
&=&(f_e+Q)_p^q\{a^p_q\}_N+\frac{1}{2}g_{pq}^{rs}\{a^{pq}_{rs}\}_N+E^{[1]}, \quad p, q, r, s \in \mbox{ PES}\label{npQEDF}
\end{eqnarray}
with $(f_e)_p^q$ and $E^{[1]}$ defined in Eqs. \eqref{Feop} and \eqref{E1full}, respectively. Here, the subscript $N$ indicates that the normal ordering
is taken with respect to $|N;\tilde{0}\rangle$. The $H_{+}^{\rm{QED}}$ Hamiltonian \eqref{npQED},
along with $\tilde{Q}$ \eqref{QVP}, $\bar{Q}$ \eqref{QSETot} and $g_{pq}^{rs}$ \eqref{Gtrans}, was already obtained
by Shabaev\cite{np-eQED} but in a top-down fashion. The aforementioned potential
dependence in the calculated energies can be removed by introducing the following correction\cite{QED1999}
\begin{eqnarray}
E^{(2)}_{\mathrm{PC}}&=&\frac{(V_{HF})_{\tilde i}^iU_i^{\tilde i}+U_{\tilde i}^i(V_{HF})_i^{\tilde i}- U_{\tilde i}^iU_i^{\tilde i}}
{\epsilon_{\tilde i}-\epsilon_{i}},\label{EPC}
\end{eqnarray}
where $U$ is the potential in Eq. \eqref{effDEQ}.
One then has a potential-independent no-pair QED (PI-QED) Hamiltonian\cite{eQED}
\begin{eqnarray}
H_+^{\mathrm{PI-QED}}&=&(f_e+Q-U)_p^q\{a^p_q\}_N+\frac{1}{2}g_{pq}^{rs}\{a^{pq}_{rs}\}_N \nonumber\\
&+& E^{[1]} + E^{(2)}_{\rm PC}, \quad p, q, r, s \in \mbox{ PES}. \label{PIQED}
\end{eqnarray}
Neglecting the $Q$ term in $H_+^{\rm{PI-QED}}$ leads to the potential-independent no-pair DCB (PI-DCB) Hamiltonian
\begin{eqnarray}
H_+^{\mathrm{PI-DCB}}&=&(f_e-U)_p^q\{a^p_q\}_N+\frac{1}{2}g_{pq}^{rs}\{a^{pq}_{rs}\}_N + E^{[1]}_{\mathrm NP} + E^{(2)}_{\mathrm PC}, \quad p, q, r, s \in \mbox{ PES}.\label{PIDCB}
\end{eqnarray}
Further neglecting the $Q$ term in $H_+^{\mathrm{PI}}$ leads to the standard no-pair DCB Hamiltonian
\begin{eqnarray}
H_{+}^{\mathrm{DCB}}=(f_e)_p^q\{a^p_q\}_N+\frac{1}{2}g_{pq}^{rs}\{a^{pq}_{rs}\}_N+E^{[1]}_{\rm np}, \quad p, q, r, s \in \mbox{ PES},\label{DCB+}
\end{eqnarray}
which has been the basis of ``no-pair relativistic quantum chemistry''.
\subsection{Quasi-four-component}
The previous no-pair four-component approaches first generate both PES and NES at the mean-field level but then discard the NES
at a correlated level.
The question is how to avoid the NES from the outset. Actually, this can be done in two different ways.
One is to retain the aesthetically simple four-component structure but \emph{freeze} the NES, while
the other is to \emph{remove} the NES so as to obtain a two-component approach.
While the former employs the untransformed Hamiltonian and
introduces approximations from the very beginning, the latter invokes an effective
Hamiltonian and has to introduce suitable approximations at a later stage.
Note that in each case the approximations introduced to the Hamiltonians
are orders of magnitude smaller than other sources of errors (e.g., incompleteness in the one- and many-particle bases)
and are therefore hardly ``approximate''. Moreover, since the two paradigms stem from precisely the same physics, they should be made fully equivalent.

\begin{figure}[t]
\centering
\begin{center}
\includegraphics[width=0.6\columnwidth,keepaspectratio=true]{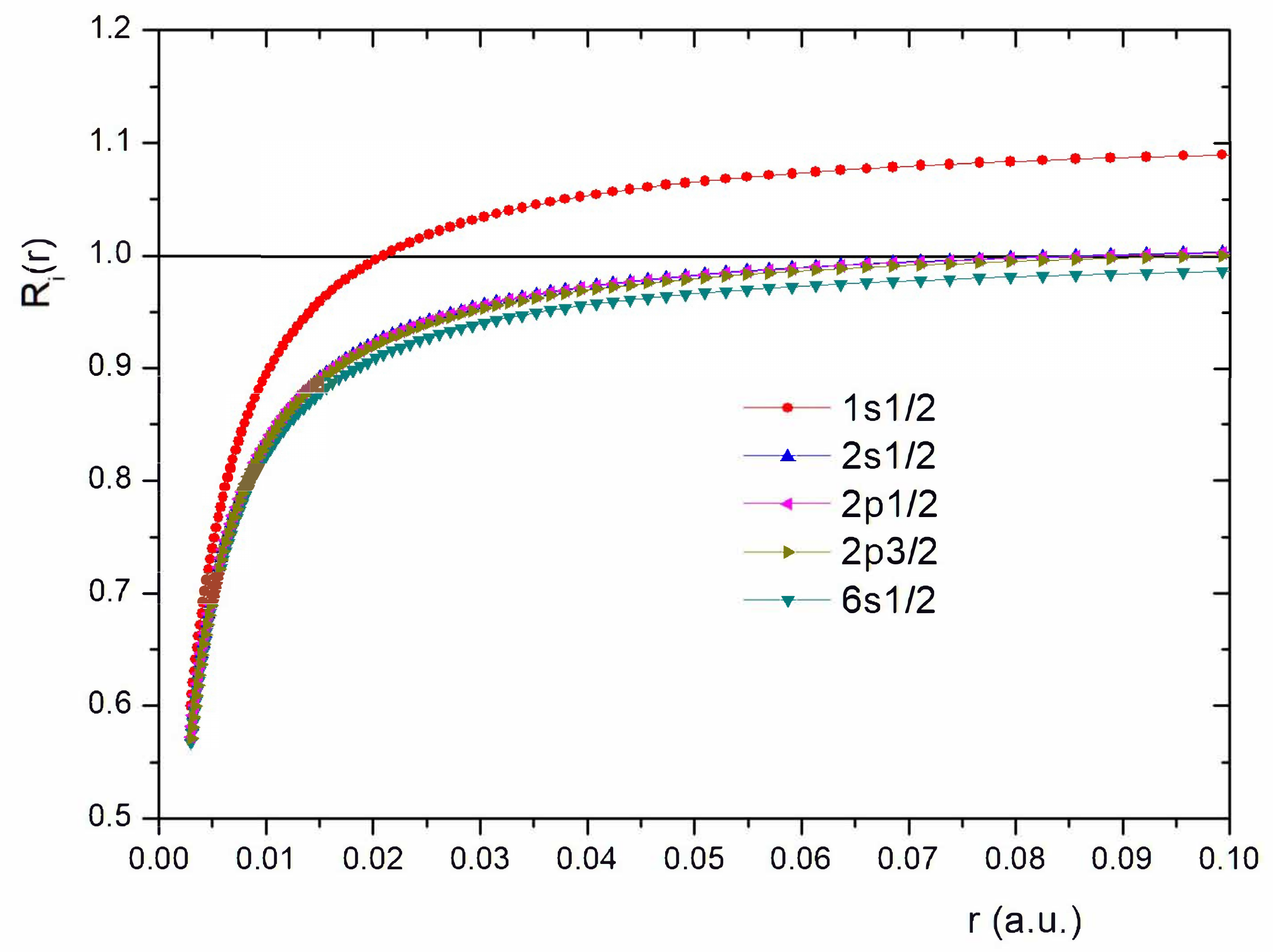}
\caption{The $R_i(r)$ operator \eqref{Rop} with $V=V_{N}+V_{H}+V_{LDA}$ as a function of the distance from the position of Rn. The radial expectation values of
$1s_{1/2}$, $2s_{1/2}$, $2p_{1/2}$, $2p_{3/2}$, and $3s_{1/2}$ are 0.015, 0.063, 0.051, 0.060, and 0.163 a.u.,
respectively. 
}
\label{FigureRop}
\end{center}
\end{figure}

To realize the first paradigm, we first take a look at the $S/L$ ratio between the small and large components of a PES $\psi_i$,
which can be obtained from the second row of Eq. \eqref{effDEQ}
\begin{eqnarray}
\psi_i^S&=&\frac{\alpha}{2}R_i\boldsymbol{
\sigma}\cdot\boldsymbol{p}\psi_i^L,\label{ExactKB}\\
R_i(\boldsymbol{r})&=&[1+\frac{\alpha^2}{2}(\epsilon_i-V(\boldsymbol{r}))]^{-1}
 \stackrel{\alpha\rightarrow 0}\rightarrow 1.\label{Rop}
\end{eqnarray}
The major effect of $\boldsymbol{\sigma}\cdot\boldsymbol{p}$ is to change the parity of the large component to that of the small component.
So the $S/L$ ratio is determined mainly by the $R_i(\boldsymbol{r})$ operator \eqref{Rop}. As can be seen from Fig. \ref{FigureRop},
the effect of $R_i(\boldsymbol{r})$ is extremely short ranged: each $R_i(\boldsymbol{r})$ becomes just
a constant factor beyond a small radius $r_c$ (ca. $0.05$ a.u.,
roughly the radii of $2s$ and $2p$). Imagine we have first solved
the (radial) Dirac equation for each isolated
(spherical and unpolarized) atom and thus obtained the
corresponding atomic 4-spinors (A4S) $\{\varphi_{\mu}\}$. Then, the atoms are
brought together to synthesize the molecule. While both the large and small
components of $\varphi_{\mu}$ will change, the $S/L$ ratio will \emph{not}!\cite{Q4CX2C,LCA4S}.
The mathematical realization\cite{BDF1} of such a physical picture is to expand
the molecular 4-spinors (M4S) $\psi_i$ in the basis only of \emph{positive-energy} A4S $\{|\varphi_{+,\mu}\rangle\}$, viz.,
\begin{align}
|\psi_i\rangle=\sum_{\mu}|\varphi_{{+,\mu}}\rangle C_{\mu i}=\sum_{\mu}\begin{pmatrix}|\varphi_{+,\mu}^L\rangle\\ |\varphi_{+,\mu}^S\rangle\end{pmatrix}C_{\mu i},\label{LCA4S}
\end{align}
which gives rise to the following projected four-component (P4C) approach\cite{BDF1}
\begin{eqnarray}
\mathbf{h}_+^{\mathrm{P4C}}\mathbf{C}&=&\mathbf{S}^{\mathrm{P4C}}_+\mathbf{C}\boldsymbol{\epsilon},\\
(h^{\mathrm{P4C}}_+)_{\mu\nu}&=&\langle \varphi^L_{+,\mu}|V|\varphi^L_{+,\nu}\rangle+\langle\varphi^S_{+,\mu}|c\boldsymbol{\sigma}\cdot\boldsymbol{p}|\varphi^L_{+,\nu}\rangle\nonumber\\
&+&\langle\varphi^L_{+,\mu}|c\boldsymbol{\sigma}\cdot\boldsymbol{p}|\varphi^S_{+,\nu}\rangle+\langle\varphi^S_{+,\mu}|V-2mc^2|\varphi^S_{+,\nu}\rangle,\label{HP4C}\\
(S^{\mathrm{P4C}}_+)_{\mu\nu}&=&\langle \varphi^L_{+,\mu}|\varphi^L_{+,\nu}\rangle+\langle \varphi^S_{+,\mu}|\varphi^S_{+,\nu}\rangle.\label{SP4C}
\end{eqnarray}
The dimension of $\mathbf{h}_+^{\mathrm{P4C}}$ is $2N^L$ instead of $4N^L$. That is, the molecular NES are excluded completely.
Physically, this amounts to neglecting rotations between the PES and NES of the isolated atoms, a kind of polarization on the atomic vacua induced
by the molecular field. As molecular formation is a very-low energy process,
its $\mathcal{O}(c^{-4})$ perturbation on the vacuum introduces no discernible errors at all\cite{BDF1,Q4C,Q4CX2C}. By further introducing
a ``model small component approximation'' (MSCA), a quasi-four-component (Q4C) approach\cite{Q4C} can be obtained, which is four-component in structure
but is computationally very much like a two-component approach. Without going into further details (see Refs. \citenum{LiuMP} and \citenum{X2CBook2017}
for the matrix elements $f^{\mathrm{Q4C}}_{pq}$ of $f_+^{\mathrm{Q4C}}$), we now have the following second-quantized, normal-ordered many-electron Hamiltonian
\begin{eqnarray}
H^{\mathrm{Q4C}}_+=E_{\mathrm{ref}}^{\mathrm{Q4C}}+f^{\mathrm{Q4C}}_{pq}\{a^p_q\}+\frac{1}{2}g_{pq}^{rs}\{a^{pq}_{rs}\}.\label{HQ4C}
\end{eqnarray}
Q4C shares precisely the same integral transformation and correlation treatment as two-component approaches\cite{LiuMP,X2CBook2017},
but does not suffer from picture-change errors (PCE)\cite{PCE} which otherwise plague two-component approaches. Moreover,
the model spectral form \cite{ShabaevModelSE,ShabaevModelSEcode2018} of the $Q$ potential \eqref{QDef} can readily be incorporated into $f^{\mathrm{Q4C}}_{pq}$, thereby leading to an QED$@$Q4C approach.

\subsection{Two-component}
By definition, a two-component relativistic theory is to transform away the positronic degrees of freedom of the Dirac operator,
so as to obtain a Hamiltonian that describes only electrons. This can be done with either unitary transformation or
elimination of the small component (ESC). However, neither route
can be done in closed form, except for the trivial free-particle
case. As such, only approximate two-component (A2C) operator (analytic)
Hamiltonians such as the Breit-Pauli Hamiltonian and the
zeroth-order regular approximation (ZORA)\cite{CPD,ZORA1} can be obtained in this way.
The situation is changed dramatically when going to the matrix formulation, where the exact decoupling
is readily achieved. In essence, the matrix formulation amounts to block-diagonalizing the matrix Dirac equation \eqref{DEQMat},
which can be done in one step\cite{X2C2005,X2C2009,X2C2007kutz,SaueX2C}, two steps\cite{BSS1,BSS2,Jensen2005} and multiple steps\cite{ReiherDKH-I,ReiherDKH-II,DKH-Peng2009}.
The three types of formulations share the same decoupling condition and differ only in the renormalization\cite{LiuMP}.
There exist even closed mapping relations among three formulations\cite{LiuMP}. Since
the initio free-particle transformation invoked in the two-step and multiple-step formulations is only necessary for finite orders\cite{DKH21986,DKH21989} but not
for infinite order, it is clear that it is the one-step formulation that should be advocated. This approach has been coined ``exact two-component'' (X2C)\cite{X2Cname}. For generality, we extend Eq. \eqref{DEQMat} to a generic eigenvalue problem
\begin{eqnarray}
\mathbf{hC}&=&\mathbf{MC} E,\label{Gendeq}\\
\mathbf{h}&=&\left(\begin{array}{cc}
\mathbf{h}_{11} & \mathbf{h}_{12} \\
\mathbf{h}_{21} & \mathbf{h}_{22} \\
\end{array}\right)=\mathbf{h}^{\dagger},\quad
\mathbf{M}=\left(\begin{array}{cc}
\mathbf{S}_{11} & \mathbf{0} \\
\mathbf{0} & \mathbf{S}_{22} \\
\end{array}\right)=\mathbf{M}^{\dagger},\quad
\mathbf{C}=\left(
\begin{array}{cc}
\mathbf{A}_+&\mathbf{A}_- \\
\mathbf{B}_+&\mathbf{B}_- \\
\end{array}\right).\label{HCSCE}
\end{eqnarray}
To decouple the PES and NES, we first introduce the following formal relations
\begin{eqnarray}
\mathbf{B}_+=\mathbf{X}\mathbf{A}_+,\quad \mathbf{A}_-= \tilde{\mathbf{X}}\mathbf{B}_-,\label{Xdef}
\end{eqnarray}
between the small- and large-component coefficients for the PES and NES, respectively.
The following unitary transformation matrix $\mathbf{U}_{X}$ can then be introduced\cite{LiuMP}
\begin{eqnarray}
\mathbf{U}_{X}&=&\boldsymbol{\Omega}_{N}\boldsymbol{\Omega}_{D},\quad
\boldsymbol{\Omega}_{N}=\left(\begin{array}{cc}
\mathbf{R}^{\dagger}_{+} & \mathbf{0} \\
\mathbf{0} & \mathbf{R}^{\dagger}_{-}
\end{array}\right),\quad
\boldsymbol{\Omega}_{D}=\left(\begin{array}{cc}
\mathbf{I} & \mathbf{X}^{\dagger} \\
\tilde{\mathbf{X}}^{\dagger} & \mathbf{I} \\
\end{array}\right),\label{Utrans2}
\end{eqnarray}
where\cite{X2C2009}
\begin{eqnarray}
\mathbf{R}_{+}&=&(\mathbf{S}^{-1}_{11}\tilde{\mathbf{S}}_{+})^{-\frac{1}{2}}=
\mathbf{S}^{-\frac{1}{2}}_{11}(\mathbf{S}^{-\frac{1}{2}}_{11}\tilde{\mathbf{S}}_{+}
\mathbf{S}^{-\frac{1}{2}}_{11})^{-\frac{1}{2}}\mathbf{S}^{\frac{1}{2}}_{11},\label{Rplus}\\
\mathbf{R}_{-}&=&(\mathbf{S}^{-1}_{22}\tilde{\mathbf{S}}_{-})^{-\frac{1}{2}}=
\mathbf{S}^{-\frac{1}{2}}_{22}(\mathbf{S}^{-\frac{1}{2}}_{22}\tilde{\mathbf{S}}_{-}
\mathbf{S}^{-\frac{1}{2}}_{22})^{-\frac{1}{2}}\mathbf{S}^{\frac{1}{2}}_{22},\label{Rminus}\\
\tilde{\mathbf{S}}_{+}&=&\mathbf{S}_{11}+\mathbf{X}^{\dagger}\mathbf{S}_{22}\mathbf{X},\\
\tilde{\mathbf{S}}_{-}&=&\mathbf{S}_{22}+\tilde{\mathbf{X}}^{\dagger}\mathbf{S}_{11}\tilde{\mathbf{X}}.
\end{eqnarray}
The requirement that $\mathbf{U}_X\mathbf{M}\mathbf{U}_X^\dag=\mathbf{M}$ leads to
\begin{eqnarray}
\tilde{\mathbf{X}}=-\mathbf{S}^{-1}_{11}\mathbf{X}^{\dagger}\mathbf{S}_{22},\label{tildeX}
\end{eqnarray}
meaning that $\tilde{\mathbf{X}}$ is determined directly by $\mathbf{X}$, which is further determined by
$(\mathbf{U}_X\mathbf{h}\mathbf{U}_X^\dag)_{21}=\mathbf{0}$, viz.,
\begin{eqnarray}
\mathbf{h}_{21}+\mathbf{h}_{22}\mathbf{X}&=&\mathbf{S}_{22}\mathbf{X}\mathbf{S}_{11}^{-1}\mathbf{L}_+^{\mathrm{UESC}},\quad \mathbf{L}_+^{\mathrm{UESC}}=\mathbf{h}_{11}+\mathbf{h}_{12}\mathbf{X},\label{XdecoupleUESC}\\
          &=&\mathbf{S}_{22}\mathbf{X}\tilde{\mathbf{S}}^{-1}_+\mathbf{L}_+^{\mathrm{NESC}}.\label{XdecoupleNESC}
\end{eqnarray}
The $\mathbf{U}_{X}$-transformation of Eq. \eqref{Gendeq} then yields
\begin{align}
&(\mathbf{U}_{X}\mathbf{h}\mathbf{U}^{\dagger}_{X})\mathbf{C}_{X}=
\left(\begin{array}{cc} \mathbf{f}_{+}^{\mathrm{X2C}} & \mathbf{0} \\
\mathbf{0} & \mathbf{f}_{-}^{\mathrm{X2C}}\\
\end{array}\right)\mathbf{C}_{X}
=\mathbf{M}\mathbf{C}_{X} E,\label{hDtrans}\\
&\mathbf{C}_{X}=(\mathbf{U}^{\dagger}_{X})^{-1}\mathbf{C}=\mathbf{M}^{-1}\mathbf{U}_{X}\mathbf{M}\mathbf{C}
=\left(\begin{array}{cc} \mathbf{C}_{+} & \mathbf{0} \\
\mathbf{0} & \mathbf{C}_{-}\\
\end{array}\right).\label{CX}
\end{align}
The upper-left block of Eq. \eqref{hDtrans} defines the equation for the PES,
\begin{eqnarray}
\mathbf{f}_{+}^{\mathrm{X2C}}\mathbf{C}_{+}&=&\mathbf{S}_{11}\mathbf{C}_{+} E_+,\label{FWeq}\\
\mathbf{f}_{+}^{\mathrm{X2C}}&=&\mathbf{R}_{+}^{\dagger}\mathbf{L}^{\mathrm{X}}_{+}\mathbf{R}_{+},\quad
\mathrm{X}=\mathrm{NESC, SESC},\label{hplus}\\
\mathbf{L}^{\mathrm{NESC}}_{+}&=&\mathbf{h}_{11}+\mathbf{h}_{12}\mathbf{X}
+\mathbf{X}^{\dagger}\mathbf{h}_{21}+\mathbf{X}^{\dagger}\mathbf{h}_{22}\mathbf{X},\label{NESC}\\
\mathbf{L}^{\mathrm{SESC}}_{+}&=&\frac{1}{2}(\tilde{\mathbf{S}}_{+}\mathbf{S}^{-1}_{11}\mathbf{L}^{\mathrm{UESC}}_{+}+c.c.),\label{SESC}\\
\mathbf{C}_{+}&=&\mathbf{R}^{-1}_{+}\mathbf{A}_{+},\label{cplus}
\end{eqnarray}
Here the acronyms UESC, NESC and SESC refer to the unnormalized, normalized\cite{NESC} and symmetrized\cite{Q4CX2C}
eliminations of the small component, respectively. Eq. \eqref{XdecoupleNESC} arises from Eq. \eqref{XdecoupleUESC} via
the relation $\mathbf{S}_{11}^{-1}\mathbf{L}_+^{\mathrm{UESC}}=\tilde{\mathbf{S}}^{-1}_+\mathbf{L}_+^{\mathrm{NESC}}$
(because $\mathbf{L}_+^{\mathrm{UESC}}\mathbf{A}_+=\mathbf{S}_{11}\mathbf{A}_+E_+$ and $\mathbf{L}_+^{\mathrm{NESC}}\mathbf{A}_+=\tilde{\mathbf{S}}_{+}\mathbf{A}_+E_+$), whereas Eq. \eqref{SESC}
arises from $\mathbf{L}^{\mathrm{SESC}}_{+}=\frac{1}{2}(\mathbf{L}^{\mathrm{NESC}}_{+}+\mathbf{L}^{\mathrm{NESC}}_{+})$
 and the decoupling condition \eqref{XdecoupleUESC}.
Likewise, the lower-right block of Eq. \eqref{hDtrans} defines the equation for the NES,
\begin{eqnarray}
\mathbf{f}_{-}^{\mathrm{X2C}}\mathbf{C}_{-}&=&\mathbf{S}_{22}\mathbf{C}_{-} E_-,\\
\mathbf{f}_{-}^{\mathrm{X2C}}&=&\mathbf{R}_{-}^{\dagger}\mathbf{L}^{\mathrm{X}}_{-}\mathbf{R}_{-},\quad\mathrm{X}=\mathrm{NESC, SESC},\label{hminus}\\
\mathbf{L}^{\mathrm{NESC}}_{-}&=&\mathbf{h}_{22}+\mathbf{h}_{21}\tilde{\mathbf{X}}+
\tilde{\mathbf{X}}^{\dagger}\mathbf{h}_{12}+\tilde{\mathbf{X}}^{\dagger}\mathbf{h}_{11}\tilde{\mathbf{X}},\label{NESCminus}\\
\mathbf{L}^{\mathrm{SESC}}_{-}&=&\frac{1}{2}(\tilde{\mathbf{S}}_{-}\mathbf{S}^{-1}_{22}\mathbf{L}^{\mathrm{UESC}}_{-}+c.c.),
\quad
\mathbf{L}^{\mathrm{UESC}}_{-}=\mathbf{h}_{22}+\mathbf{h}_{21}\tilde{\mathbf{X}},\label{SESCminus}\\
\mathbf{C}_{-}&=&\mathbf{R}^{-1}_{-}\mathbf{B}_{-}.\label{cminus}
\end{eqnarray}
It can be proven\cite{X2CSOC2} that $\mathbf{C}_+$ ($\mathbf{C}_-$) is closest to $\mathbf{A}_+$ ($\mathbf{B}_-$) in the least-squares sense.

The above manipulation can further be extended to include magnetic fields as well\cite{PhysRep,X2CBook2017}.
Moreover, at variance with the explicit expression \eqref{hplus}, $\mathbf{f}_+^{\mathrm{X2C}}$  can also be constructed on the fly,
by an orthonormalization and back-transformation procedure\cite{SaueX2C}.
The following remarks are still in order.
\begin{enumerate}[(1)]
\item The one-step matrix formulation of two-component relativistic theories was initiated by Dyall\cite{NESC} in 1997. However, the proper formulation of the (energy-independent) decoupling condition \eqref{XdecoupleUESC}/\eqref{XdecoupleNESC}\cite{X2C2005}
as well as the correct renormalization \eqref{Rplus}\cite{X2C2009} were found only later on.
It was also found\cite{LiuMP} that the same results can be obtained by converting
the Foldy-Wouthuysen (FW) Hamiltonian \cite{FW1950} (which has no closed form though)
directly into matrix form in terms of the RKB basis. That is, the matrix and operator (more precisely operator-like) formulations of X2C are identical, as should be.
The situation is different for finite-order A2C approaches. To see this, we look at the ZORA equation\cite{CPD,ZORA1},
\begin{eqnarray}
(V+T^{\mathrm{ZORA}})\psi_p^{\mathrm{ZORA}}&=&\psi_p^{\mathrm{ZORA}}\boldsymbol{\epsilon}_p^{\mathrm{ZORA}},\label{ZORA}\\
T^{\mathrm{ZORA}}&=&\boldsymbol{\sigma}\cdot\boldsymbol{p}\frac{1}{2-\alpha^2V}\boldsymbol{\sigma}\cdot\boldsymbol{p}.
\end{eqnarray}
In view of the identity $1/(2-\alpha^2V)\times (2-\alpha^2V) =1 $, the matrix elements of $T^{\mathrm{ZORA}}$ can be calculated as
\begin{eqnarray}
\langle\boldsymbol{\sigma}\cdot\boldsymbol{p}g_{\mu}|\frac{1}{2-\alpha^2V}|\boldsymbol{\sigma}\cdot\boldsymbol{p}g_{\rho}\rangle
[(2\mathbf{T})^{-1}]_{\rho\sigma}\langle \boldsymbol{\sigma}\cdot\boldsymbol{p}g_{\sigma}|2-\alpha^2V|\boldsymbol{\sigma}\cdot\boldsymbol{p}g_{\nu}\rangle=2\mathbf{T}_{\mu\nu},\label{RKBidentity}
\end{eqnarray}
which leads to
\begin{eqnarray}
\mathbf{T}^{\mathrm{ZORA}}=\mathbf{T}\mathbf{X}^{\mathrm{ZORA}},\quad \mathbf{X}^{\mathrm{ZORA}}=(\mathbf{T}-\frac{\alpha^2}{4}\mathbf{W})^{-1}\mathbf{T}.\label{ZORAXmat}
\end{eqnarray}
Therefore, the matrix representation of the ZORA equation \eqref{ZORA} reads
\begin{eqnarray}
(\mathbf{V}+\mathbf{T}\mathbf{X}^{\mathrm{ZORA}})\mathbf{A}^{\mathrm{ZORA}}=\mathbf{S}\mathbf{A}^{\mathrm{ZORA}}\boldsymbol{\epsilon}^{\mathrm{ZORA}}.\label{ZORAMat}
\end{eqnarray}
Thanks to the use of the resolution of the identity (RI) in terms of the $\{\boldsymbol{\sigma}\cdot\boldsymbol{p}g_{\mu}\}$ basis in Eq. \eqref{RKBidentity},
the matrix ZORA equation \eqref{ZORAMat} agrees with the operator ZORA equation \eqref{ZORA} only when the basis $\{g_{\mu}\}$ is complete.
This is totally different from the matrix counterpart (i.e., X2C) of the (non-expanded) FW Hamiltonian \cite{FW1950}, where the use
of the same RI is \emph{not} an approximation but only a formal step\cite{LiuMP}. Since Eq. \eqref{ZORAMat}
is never used in practice, the commonly called ZORA (and the infinite-order regular approximation\cite{IORA}) is a
genuine analytic relativistic theory. In contrast, other relativistic theories, whether finite-order\cite{DKH21986,DKH21989} or infinite-order\cite{X2C2005,X2C2009,X2C2007kutz,SaueX2C,BSS1,BSS2,Jensen2005,ReiherDKH-I,ReiherDKH-II,DKH-Peng2009},
are all algebraic. However, the analyticity of a relativistic Hamiltonian should not be celebrated, simply because only
Fock space is the right framework for relativistic quantum mechanics,
which gives rise to only algebraic relativistic many-electron Hamiltonians (see Sec. \ref{SeceQED}).

\item The eigenvalue equation \eqref{FWeq} and the decoupling condition \eqref{XdecoupleUESC}/\eqref{XdecoupleNESC} are coupled
and have to be solved iteratively\cite{X2C2005,X2C2007kutz}. The so-obtained results agree with those by the parent matrix Dirac equation \eqref{DEQMat}
up to machine accuracy, thereby justifying the name ``exact two-component''\cite{X2Cname}. However, the computation is much more expensive
than solving Eq. \eqref{DEQMat} directly, even for a one-electron system. Therefore, a suitable approximation to $\mathbf{X}$ must be found in order to
make X2C practical. To this end, we take a look at the matrix presentation of the
key relation \eqref{ExactKB} in a RKB basis \eqref{RKB} (without caring for the inherent singularities\cite{X2C2006kutz}), viz.,
\begin{eqnarray}
\mathbf{B}_{+,i}&=&\frac{1}{2}\mathbf{T}^{-1}\mathbf{R}^{(i)}\mathbf{A}_{+,i},\quad \mathbf{R}^{(i)}_{\mu\nu}=\langle g_{\mu}|\boldsymbol{\sigma}\cdot\boldsymbol{p}R_i\boldsymbol{\sigma}\cdot\boldsymbol{p}|g_{\nu}\rangle,\\
&=&\mathbf{U}^{(i)}\mathbf{A}_{+,i},
\end{eqnarray}
where $\mathbf{U}^{(i)}$ is the energy/state-dependent equivalent\cite{NESC} of the state-universal $\mathbf{X}$.
Since the $R_i(\boldsymbol{r})$ operator is extremely shorted ranged (see Fig. \eqref{FigureRop}), it can be envisaged
that the molecular $\mathbf{U}^{(i)}$ (and hence $\mathbf{X}$) should be strongly block-diagonal in atoms.
As can be seen from Fig. \ref{FigureX}, this is indeed the case. Note in particular that, to enhance the interatomic
interaction, we have set the interatomic distance of \ce{Au2} to 1.5 \AA, which is much shorter than the equilibrium distance of 2.47~\AA.
Therefore, a general deduction is that
the molecular $\mathbf{X}$ can well be approximated as the superposition of the atomic ones\cite{Q4C,Q4CX2C,LCA4S,Dyall01}
\begin{eqnarray}
\mathbf{X}=\sum_F^{\oplus}\mathbf{X}_F,\label{AtomX}
\end{eqnarray}
which stays in the same spirit as P4C (see Point \ref{P4C2NESC} below). Here,
each atomic $\mathbf{X}_F$ can, in view of the very definition \eqref{Xdef},
be obtained by solving the (radial) Dirac equation for a neutral or ionic spherical and unpolarized configuration.
The atomic approximation to $\mathbf{X}$ works very well not only for ground state energies of molecular systems\cite{Q4C,Q4CX2C}, but also for electric\cite{X2CTDReO4,X2CTDYbO,X2CTDOsO4,X2CTDSOC,X2CEOMCC,RTDDFTrev} and magnetic\cite{X2CNMR2009,X2CNMR2012} response properties, analytic energy gradients and Hessian\cite{X2Cgrd2020}, as well as periodic systems\cite{X2C-PBC}.
In contrast, the widely used approximation $\mathbf{X}_{1e}$ obtained by diagonalizing the one-electron Dirac matrix is
not accurate enough for nuclear magnetic shielding, and cannot be applied to periodic systems. There have been attempts\cite{DLU,Nakai2012a,Nakai2012b} to approximate
the renormalization matrix $\mathbf{R}_+$ \eqref{Rplus} also as the superposition of the atomic ones. Since $\mathbf{R}_+$
is much less local than $\mathbf{X}$, such approximation does introduce discernible errors\cite{LiuGVVPT2}.
Nevertheless, such errors are tolerable for large systems in view of the dramatic gain in computational efficiency (especially in
gradient and Hessian calculations\cite{X2Cgrd2020}). The atomic approximation to $\mathbf{X}$ (and $\mathbf{R}_+$) can obviously be generalized
to a diatomic (fragmental) approximation\cite{Q4C,Q4CX2C}, which is of course only necessary if one is interested in highly distorted
molecular systems in which two heavy atoms are very close to each other in distance.
It is of particular interest to note that
the atomic/fragmental approximation to both $\mathbf{X}$ and $\mathbf{R}_+$ (i.e., the X2C/AU Ansatz defined in Ref. \citenum{X2Cgrd2020})
allows one to interpret\cite{LiuMP} X2C as a seamless bridge between the Dirac and Schr{\"o}dinger equations, because it can treat the heavy and light atoms
in the system relativistically and nonrelativistically, respectively, unlike that the Dirac (Schr{\"o}dinger) equation treats
the whole system relativistically (nonrelativistically).
\begin{figure}
\includegraphics[width=0.35\textwidth]{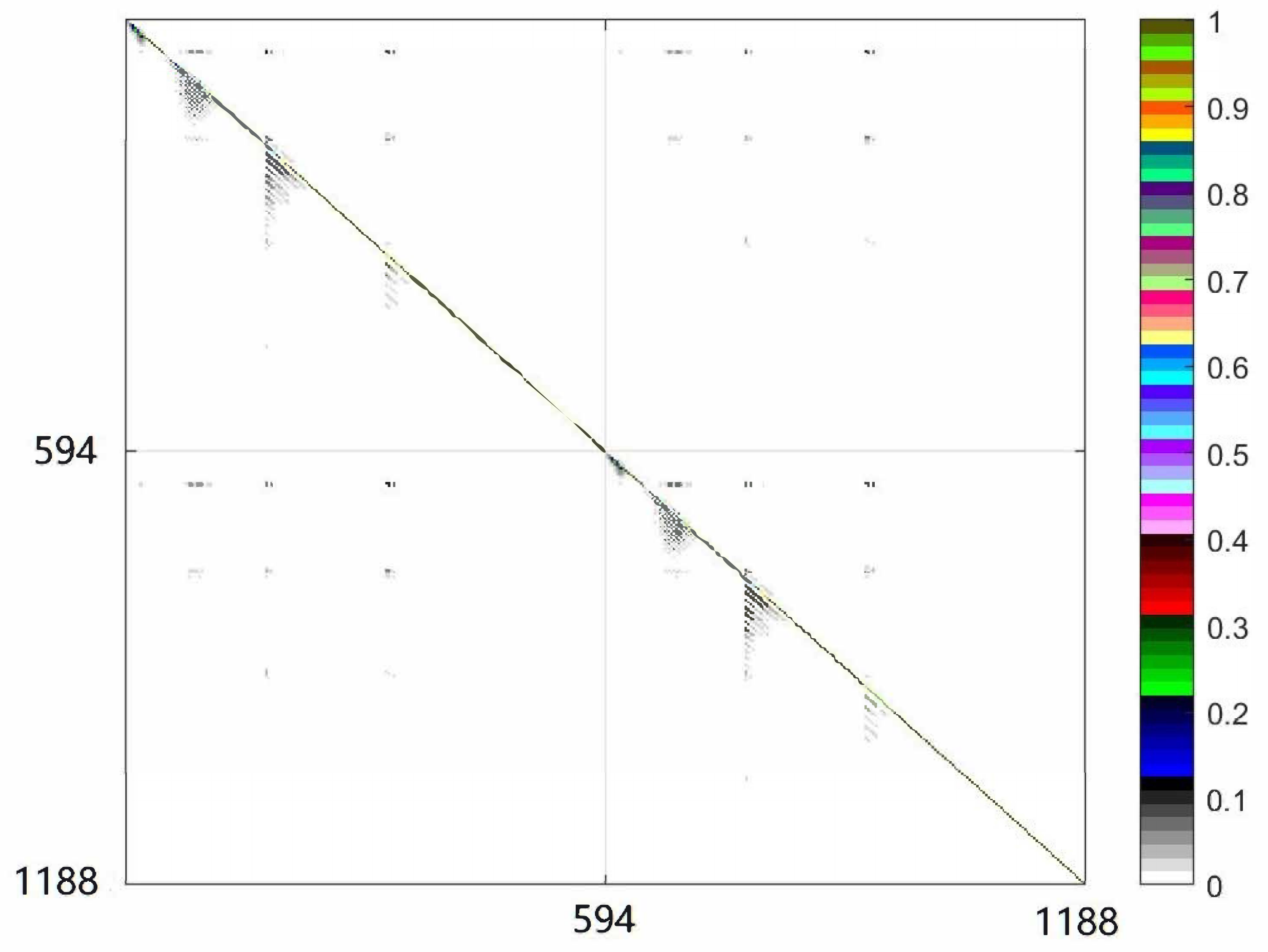}
\caption{Distribution of the matrix elements of $\mathbf{X}=\mathbf{B}_+\mathbf{A}_+^\dag(\mathbf{A}_+\mathbf{A}_+^\dag)^{-1}$ \eqref{Xdef}
for \ce{Au2} at a distance of 1.5~\AA. Dirac-Hartree-Fock (DHF) result with the uncontracted ANO-RCC basis set (594 functions for each atom).}
\label{FigureX}
\end{figure}
\item\label{P4C2NESC} $\mathbf{L}^{\mathrm{NESC}}_{+}$ \eqref{NESC} is closely related to $\mathbf{h}_+^{\mathrm{P4C}}$ \eqref{HP4C}. To see this, we assume the A4S $\{|\varphi_{+,\mu}\rangle\}$ in the latter are further expanded in a RKB basis, viz.,
\begin{eqnarray}
|\varphi_{+,\mu}\rangle=\sum_{\lambda\in K} \begin{pmatrix} g_{\lambda}a_{K,\lambda\mu}\\ \frac{\alpha}{2}\boldsymbol{\sigma}\cdot\boldsymbol{p}g_{\lambda}b_{K,\lambda\mu}\end{pmatrix},\quad \forall \mu \in K; \quad \mathbf{b}_K=\mathbf{X}_K\mathbf{a}_K
\end{eqnarray}
for each atom $K$. We then have
\begin{eqnarray}
\mathbf{h}_+^{\mathrm{P4C}}&=&\mathbf{a}^\dag \mathbf{L}^{\mathrm{NESC}}_{+} \mathbf{a}, \quad \mathbf{a}=\sum_{K}^{\oplus}\mathbf{a}_K, \label{HP4C-RKB}\\
\mathbf{S}_+^{\mathrm{P4C}}&=&\mathbf{a}^\dag\tilde{\mathbf{S}}_+\mathbf{a},\label{SP4C-RKB}
\end{eqnarray}
where $\mathbf{L}^{\mathrm{NESC}}_{+}$ in Eq. \eqref{HP4C-RKB} and $\tilde{\mathbf{S}}_+$ in Eq. \eqref{SP4C-RKB}
have adopted the atomic approximation \eqref{AtomX} to $\mathbf{X}$. It is hence clear that P4C
is just NESC, provided that the atomic-natural-spinor-type generally
contracted RKB basis and the atomic approximation \eqref{AtomX} to $\mathbf{X}$ are used in both cases.
However, P4C\cite{BDF1} and NESC\cite{NESC} were introduced in completely different ways, in the same year though. Unlike NESC,
P4C is not limited to the RKB condition. Rather, it can also adopt, e.g., numerical A4S.
\item All physical operators are subject to the same transformation going from the Dirac equation to a two-component theory.
Neglecting this will result in PCE\cite{PCE}. This can readily be done in the case of X2C, thanks to the simple relations $\mathbf{A}_+=\mathbf{R}_{+}\mathbf{C}_+$ and $\mathbf{B}_+=\mathbf{X}\mathbf{A}_+$.
Moreover, the MSCA  (which takes care of both scalar and spin-orbit one-centered two-electron picture-change effects) underlying Q4C \cite{Q4C} can also be employed in X2C.
A more dramatic simplification of X2C is to assemble the two-electron spin-orbit part of $f^{\mathrm{X2C}}_+$ from DHF calculations of spherically averaged atomic configurations and then neglect all molecular relativistic two-electron integrals\cite{SOX2CAMF}. All in all,
the second-quantized, normal-ordered, PCE-corrected many-electron X2C Hamiltonian can be written as\cite{Q4CX2C}
\begin{eqnarray}
H^{\mathrm{X2C}}_+=E_{\mathrm{ref}}^{\mathrm{X2C}}+f^{\mathrm{X2C}}_{pq}\{a^p_q\}+\frac{1}{2}g_{pq}^{rs}\{a^{pq}_{rs}\}.\label{HX2C}
\end{eqnarray}
Note in passing that if the model spectral form \cite{ShabaevModelSE,ShabaevModelSEcode2018} of the $Q$ potential \eqref{QDef}
is included in $\mathbf{h}$ \eqref{HCSCE}, we would obtain automatically an QED$@$X2C approach\cite{PhysRep}.
\end{enumerate}
At this stage it should have been clear that the Q4C and X2C formalisms render no-pair four- and two-component relativistic calculations completely identical in all aspects of simplicity,
accuracy and efficiency, at both the mean-field and correlated levels (a point that was observed more than a decade ago\cite{Q4CX2C}).
\subsection{Spin-separated two-component}
There are various situations where one would like to treat spin-free (sf) and spin-dependent (sd) relativistic effects separately.
For instance, the terms ``intersystem crossing'' and ``multistate reaction'' are both rooted in the
perturbative treatment of spin-orbit coupling (SOC).
In addition to SOC, spin-dependent interactions include also spin-spin coupling (SSC). While SOC contains both one- and two-body terms,
SSC is purely a two-body operator arising from the spin separation of the Gaunt interaction\cite{MCSCF-DPT1}
and should be taken into account in calculations of magnetic properties\cite{VahtrasSSC,NeeseSSC}.
Here we outline briefly how to extract SOC from the X2C Hamiltonian. The very first issue lies in that
$\mathbf{f}_+^{\mathrm{X2C}}$ \eqref{hplus} is defined only in matrix form, such that the Dirac identity
\begin{eqnarray}
(\boldsymbol{\sigma}\cdot\boldsymbol{A})B(\boldsymbol{\sigma}\cdot\boldsymbol{C})=\boldsymbol{A}\cdot(B\boldsymbol{C})+\mathbbm{i}\boldsymbol{\sigma}\cdot[\boldsymbol{A}\times(B\boldsymbol{C})]\label{Dirac}
\end{eqnarray}
cannot be used. However, we can start with the partitioning of the Dirac matrix \eqref{DEQMat} into a scalar and a spin-orbit term
\begin{eqnarray}
\begin{pmatrix}
\mathbf{V} & \mathbf{T} \\ \mathbf{T} & \frac{\alpha^2}{4}\mathbf{W}-\mathbf{T}\end{pmatrix}
&=&\begin{pmatrix} \mathbf{V} & \mathbf{T} \\ \mathbf{T} & \frac{\alpha^2}{4}\mathbf{W}_{sf}-\mathbf{T} \end{pmatrix}
+\begin{pmatrix}\mathbf{0} & \mathbf{0} \\ \mathbf{0} & \frac{\alpha^2}{4}\mathbf{W}_{sd} \end{pmatrix},\label{part2}
\end{eqnarray}
where
\begin{eqnarray}
(W_{sf})_{\mu\nu}=\langle g_\mu|\boldsymbol{p}\cdot V \boldsymbol{p}|g_\nu\rangle,\quad (W_{sd})_{\mu\nu}=\langle g_\mu| i\boldsymbol{\sigma}\cdot(\boldsymbol{p}V\times\boldsymbol{p}) |g_\nu\rangle.\label{WsdMat}
\end{eqnarray}
The first, spin-free term can be block-diagonalized in the same way as before, so as to obtain
\begin{eqnarray}
\mathbf{h}^{\mathrm{X2C}}_{+,sf}=\mathbf{R}_{+,0}^\dagger(\mathbf{V}+\mathbf{T}\mathbf{X}_0+\mathbf{X}_0^\dagger\mathbf{T}
+\mathbf{X}_0^\dagger[\frac{\alpha^2}{4}\mathbf{W}_{sf}-\mathbf{T}]\mathbf{X}_0)\mathbf{R}_{+,0},
\end{eqnarray}
where $p,q,r,s$ refer to real-valued spin orbitals.
Allying the spin-free $\mathbf{U}_0$ transformation [cf. Eq. \eqref{Utrans2}] to the second term of Eq. \eqref{part2} leads to
\begin{eqnarray}
\begin{pmatrix}\frac{\alpha^2}{4}\mathbf{R}_{+,0}^{\dagger}\mathbf{X}^{\dagger}_{0}\mathbf{W}_{sd}\mathbf{X}_{0}\mathbf{R}_{+,0}
&\frac{\alpha^2}{4}\mathbf{R}_{+,0}^{\dagger}\mathbf{X}^{\dagger}_{0}\mathbf{W}_{sd}\mathbf{R}_{-,0}\\ \frac{\alpha^2}{4}\mathbf{R}_{-,0}^{\dagger}\mathbf{W}_{sd}\mathbf{X}_{0}\mathbf{R}_{+,0}&\frac{\alpha^2}{4}\mathbf{R}_{-,0}^{\dagger}\mathbf{W}_{sd}\mathbf{R}_{-,0}\end{pmatrix},\label{SO1Mat}
\end{eqnarray}
where the upper-left block is just the first-order SOC (to be denoted as so-DKH1)
\begin{eqnarray}
\mathbf{h}_{SO,1e}^{(1)}=\frac{\alpha^2}{4}\mathbf{R}_{+,0}^{\dagger}\mathbf{X}_0^\dagger\mathbf{W}_{SO}\mathbf{X}_0\mathbf{R}_{+,0}.\label{hSO}
\end{eqnarray}
Higher-order SOC can readily be obtained\cite{X2CSOC1} by carrying out further DKH-type unitary transformations that eliminate at each step the lowest-order odd terms in $\mathbf{W}_{sd}$. In particular, the so-DKH2 and so-DKH3 operators $\mathbf{h}_{SO,1e}^{(n)}$
can be constructed essentially for free (see Ref. \citenum{X2CSOC2} for the explicit expressions), because
all necessary quantities are already available after constructing $\mathbf{h}^{\mathrm{X2C}}_{+,sf}$. As for the two-electron SOC,
a mean-field approximation to the first-order terms is sufficent\cite{X2CSOC2}
\begin{eqnarray}
\mathbf{f}_{SO,2e}^{(1)}&=&\frac{\alpha^2}{4}\mathbf{R}_{+,0}^{\dagger}[\mathbf{G}^{LL}_{SO}+\mathbf{G}^{LS}_{SO}\mathbf{X}_{0}+
\mathbf{X}^{\dagger}_{0}\mathbf{G}^{SL}_{SO}+\mathbf{X}^{\dagger}_{0}\mathbf{G}^{SS}_{SO}\mathbf{X}_{0}]\mathbf{R}_{+,0},\label{FockSO}
\end{eqnarray}
where $\mathbf{G}^{XY}_{SO}$ ($X, Y \in\{ L, S \}$) are the matrices of the effective one-electron operators
$G^{XY}_{SO}$,
\begin{eqnarray}
G^{XY}_{SO}&=&\mathbbm{i}\boldsymbol{\sigma}\cdot\boldsymbol{g}^{XY}=\mathbbm{i}\sum_{l}\sigma_{l}g^{XY,l},\quad X,Y\in\{L,S\},\;l\in\{x,y,z\},\label{gSO}\\
g^{LL,l}_{\mu\nu}&=&-\sum_{\lambda\kappa}2K^{l}_{\lambda\mu,\kappa\nu}P^{SS}_{\lambda\kappa},\label{gll}\\
g^{LS,l}_{\mu\nu}&=&-\sum_{\lambda\kappa}(K^l_{\mu\lambda,\kappa\nu}+K^l_{\lambda\mu,\kappa\nu})P_{\lambda\kappa}^{LS},\label{gls}\\
g^{SS,l}_{\mu\nu}&=&-\sum_{\lambda\kappa}2(K^l_{\mu\nu,\kappa\lambda}+K^l_{\mu\nu,\lambda\kappa}-K^l_{\mu\lambda,\nu\kappa})
P^{LL}_{\lambda\kappa},\label{gss}\\
K^{l}_{\mu\nu,\kappa\lambda}&=&\sum_{mn}\varepsilon_{lmn}(\mu_m\nu|\kappa_n\lambda)=-K^{l}_{\kappa\lambda,\mu\nu},\quad\mu_m=\partial_m\mu,\;
l,m,n\in\{x,y,z\},\label{SOintegrals}\\
\mathbf{P}^{LL}&=&\mathbf{R}_{+,0}\mathbf{P}\mathbf{R}_{+,0}^{\dagger},\quad
\mathbf{P}^{LS}=\mathbf{P}^{LL}\mathbf{X}^{\dagger}_0,\quad\mathbf{P}^{SS}=\mathbf{X}_{0}\mathbf{P}^{LL}\mathbf{X}^{\dagger}_0.
\end{eqnarray}
Here, $\kappa,\lambda,\mu,\nu$ refer to atomic (Gaussian) spin orbitals, $\varepsilon_{lmn}$ is the Levi-Civita symbol, while
$\mathbf{P}=\frac{1}{2}(\mathbf{P}^\alpha+\mathbf{P}^\beta)$ is the spin-averaged molecular density matrix,
with $\mathbf{P}^\alpha$ and $\mathbf{P}^\beta$ being the converged sf-X2C-ROHF/ROKS (restricted open-shell Hartree-Fock/Kohn-Sham) spin density matrices.
The terms in Eq. \eqref{gls} and the first two terms in Eq. \eqref{gss} arise from the Coulomb interaction and represent
the so-called spin-same-orbit coupling, whereas
the term \eqref{gll} and the third term of Eq. \eqref{gss} originate from
the Gaunt interaction and hence represent the spin-other-orbit coupling\cite{X2CSOC2}.
In view of the short-range nature of SOC, a one-centre approximation to
the integrals $K^{l}_{\mu\nu,\kappa\lambda}$ \eqref{SOintegrals} can further be invoked. In this case, only the atomic blocks of the
molecular density matrix $\mathbf{P}^{XY}$ contribute to $\mathbf{G}^{XY}_{SO}$. Yet, $\mathbf{f}_{SO,2e}^{(1)}$ is still a full matrix.
If wanted, the SSC\cite{MCSCF-DPT1} can readily be added to $\mathbf{f}_{SO,2e}^{(1)}$.
The second-quantized, normal-ordered, spin-separated X2C Hamiltonian then reads
\begin{eqnarray}
H^{\mathrm{X2CSOC}}_{+,n}&=&E_{\mathrm{ref}}+H_{sf}+H_{sd}^{[n]},\quad n=1 \mbox{ or } 3,\label{approximateH}\\
H_{sf}&=&[\mathbf{h}_{+,sf}^{\mathrm{X2C}}]_p^q \{a^p_q\}+\frac{1}{2} g_{pq}^{rs} \{a^{pq}_{rs}\},\label{TCHsf}\\
H_{sd}^{[n]}&=&[\mathbf{h}_{SO,1e}^{[n]}+\mathbf{f}_{SO,2e}^{(1)}]_p^q \{a^p_q\}, \label{SOCoper}
\end{eqnarray}
which is the simplest variant in the whole family of spin-separated X2C Hamiltonians\cite{X2CSOCBook2017} (NB: $[n]$ denotes up to $n$-th order).
It has been combined with both spin-adapted open-shell time-dependent density functional theory\cite{SARPA,SATDDFT,XTDDFT}
and equation-of-motion coupled cluster for calculating fine structures of electronically excited states\cite{X2CTDSOC,X2CEOMCC}.

Note in passing that, if the decoupling matrix $\mathbf{X}_{0}$ \eqref{Xdef} and the renormalization matrix $\mathbf{R}_{+,0}$ \eqref{Rplus}
are set to identity in both $\mathbf{h}_{SO,1e}^{(1)}$ \eqref{hSO} and $\mathbf{f}_{SO,2e}^{(1)}$ \eqref{FockSO} (i.e., so-DKH1), $H_{sd}^{(1)}$ \eqref{SOCoper} will reduce to the Breit-Pauli spin-orbit Hamiltonian (so-BP). So $H_{sd}^{(1)}$ can be understood as a bracketed (stabilized) so-BP. While so-BP can only used as a first-order perturbation operator on top of the nonrelativistic problem,
$H_{sd}^{[n]}$ is bounded from below and can hence be treated variationally. On the other hand,
if $\mathbf{X}_{0}$ and $\mathbf{R}_{+,0}$ in $\mathbf{h}_{SO,1e}^{(1)}$ \eqref{hSO} and $\mathbf{f}_{SO,2e}^{(1)}$ \eqref{FockSO} are both
set to the free-particle counterparts, $H_{sd}^{(1)}$ \eqref{SOCoper} will reduce to the original mean-field so-DKH1\cite{SOMF}.
Moreover, $H_{sd}^{[3]}$ is extremely accurate for both core and valence states\cite{X2CSOC1} and can therefore be regarded as an equivalent of
the non-perturbative SOX2CAMF operator\cite{SOX2CAMF}.

The various Hamiltonians discussed so far, including $H_a^{\mathrm{QED}}$ \eqref{HnH}, $H_+^{\mathrm{PI-QED}}$ \eqref{PIQED}, $H_{+}^{\rm{QED}}$ \eqref{npQED}, $H_+^{\mathrm{PI-DCB}}$ \eqref{PIDCB}, $H_+^{\mathrm{DCB}}$ \eqref{DCB+}, $H^{\mathrm{Q4C}}_+$ \eqref{HQ4C}, $H^{\mathrm{X2C}}_+$ \eqref{HX2C}, $H^{\mathrm{X2CSOC}}_{+,n}$ \eqref{approximateH},
and those A2C and nonrelativistic ones, share the same generic form
\begin{eqnarray}
H=E_{\mathrm{ref}}+f_p^q\{a^p_q\}+\frac{1}{2}g_{pq}^{rs}\{a^{pq}_{rs}\}.
\end{eqnarray}
It is just that the Fockian operator $f$ has to be interpreted differently.
A complete and continuous ``Hamiltonian Ladder'' can then be depicted\cite{IJQCrelH,PhysRep}.
The following points deserves to be emphasized again:
\begin{enumerate}[(a)]
\item Relativistic Hamiltonians can only be formulated in Fock space, whereas all
 first-quantized relativistic Hamiltonian suffer from contaminations of NES.
\item $H_a^{\mathrm{QED}}$ \eqref{HnH} is the most accurate relativistic many-electron Hamiltonian and serves as the basis of the emerging field of ``molecular QED''.
\item\label{Q4Cterm}Under the NPA, four- and two-component approaches are fully equivalent
in all aspects of simplicity, accuracy and efficiency. Therefore, one should speak of ``four-component and two-component equally good'', instead of ``four-component good, two-component bad'' or
``two-component good, four-component bad''.
\item\label{X2Cterm} X2C is computationally the same as but is much simpler and more accurate than A2C. As such, A2C should be regarded as outdated.
\item sf-X2C+so-DKH1 is computationally the same as but is more accurate than NR+so-BP. As such, NR+so-BP should be regarded as outdated.
\end{enumerate}
\section{No-pair correlation}\label{SecEc}
Having discussed extensively the QED and relativistic many-electron Hamiltonians, we comment briefly on the correlation problem.
Due to the large gap between the NES and PES, a second-order treatment of the
NES is sufficient (see Sec. \ref{SecE2}). Therefore, the major challenge still resides in
the no-pair correlation within the manifold of PES. In this regard, like the nonrelativistic case, one has to face
two general issues, i.e., the slow basis-set convergence and the strong correlation problem.
The former can only be improved by the so-called explicitly correlated methods. However,
relativistic explicit correlation is plagued by two conceptual points: (a) no-pair projected or second-quantized relativistic Hamiltonians
are simply incompatible with explicit correlation due to the lack of analytic operators. (b)
The fact\cite{KutzRelR12} that the two limits $c\rightarrow \infty$ and $r_{12}\rightarrow 0$ do not commute makes how to apply
the correlation factor $f_{12}$ (which itself is a complicated quantity\cite{RelR12}) an open question. Rather unexpectedly, although the small-component
$\psi_p^{S}$ of a PES is indeed smaller (albeit in the mean) than the large-component $\psi_p^{L}$ by a factor of $c^{-1}$,
the small-small component $\Psi^{SS}(\boldsymbol{r}_1,\boldsymbol{r}_2)$ of a two-electron wave function $\Psi(\boldsymbol{r}_1,\boldsymbol{r}_2)$ is of the \emph{same} order
of magnitude as the large-large component $\Psi^{LL}(\boldsymbol{r}_1,\boldsymbol{r}_2)$ at the coalescence point\cite{RelR12}.
This means simply that there is no obvious argument to favor the incorporation of the correlation factor $f_{12}$ in a way that is
in line with ``first $c\rightarrow\infty$ and then $r_{12}\rightarrow 0$'' or ``first $r_{12}\rightarrow 0$ and then $c\rightarrow\infty$''.
These issues have recently been scrutinized in depth\cite{RelR12,WaveFuncBook2017,CoalescenceBook2017,RelR12Book2017}.
Since there are no new numerical results thereafter, we do not repeat the discussions here. However,
it does deserve to be mentioned that the short-range density-functional type of corrections for basis-set incompleteness\cite{DFT2WFT-JCP2018,DFT2WFT-JCP2019,DFT2WFT-JPCL2019}
is highly promising, not only because of its simplicity but also because it is rooted in second quantization and is hence compatible with
relativistic Hamiltonians. The remaining issue is to develop suitable short-range relativistic density functionals for this purpose\cite{srRelDFT2020}.

Compared to the slow basis-set convergence problem, the strong correlation problem is even more intricate in practice.
A system is characterized as strongly correlated if a qualitative description already requires a multiconfigurational wave function.
The main issue here lies in that the static and dynamic components of electron correlation are often strongly entangled and even interchangeable.
The situation is further worsened by SOC. Although a number of relativistic schemes have been developed in the past\cite{RASSOC,CASSI1986,CASSI1989,CASPT2-SOC,4C-MCSCF1996,2C-CASSCF1996,2C-CASSCF2003,2C-CASSCF2013,4C-MCSCF2008,4C-CASPT22008,4C-CASSCF2015,4C-CASSCF2018,4C-icMRCI2015,RelDMRG2005,LiuGVVPT2,RelDMRG2014,RelDMRG2018,4C-MR2018,LixiaosongX2CCASSCF},
approaches that can provide a balanced and self-adaptive description of the static and dynamic components of correlation still remain to be formulated.
It is believed that the ultimate way is to introduce some selection procedure that can adapt to the variable static correlation automatically and meanwhile
can be terminated at a stage at which the residual dynamic correlation can well be described by a low-order approach.
This leads naturally to ``selected configuration interaction plus second-order perturbation theory'' (sCIPT2),
a very old idea that can be traced back to the end of 1960s and has recently been revived in various ways (see Ref. \citenum{iCIPT2} for a recent review).
Such approaches are most suited for relativistic calculations because of the following reasons:
\begin{enumerate}
\item A compact yet high-quality variational space can readily be determined by an iterative selection procedure, thereby avoiding problems\cite{iCIPT2} inherent in the scenario of
complete active space (CAS). For instance, the size of the CAS would be doubled in the presence of SOC, so as to limit severely the applicability of CAS-based
four- or two-component relativistic correlation methods.
This problem can only be resolved by selection.
\item The selection of important configurations is particularly effective for SOC, thanks to the short-range nature of SOC. This had better be combined with
a local representation\cite{FLMO1,Triad,ACR-FLMO,FLMO3} from the outset.
\item Unlike nonlinear wave function Ans\"atze, the symmetry adaptation of CI wave functions can readily be achieved by means of the
spin-dependent unitary group approach\cite{SD-UGA3,PitzerSOCI}.
\item SOC is strongly dominated by the one-body terms, such that a second-order perturbative treatment of dynamic correlation, on top of
a well-controlled variational space, should be sufficient.
\end{enumerate}
An X2C-based heat bath CI version\cite{HBCISOC} of sCIPT2 has just been realized, showing great promises
although SOC is included therein only at the correlated level but not at the orbital level. The combination of the
QED$@$Q4C or QED$@$X2C Hamiltonian with the
recently proposed iCIPT2\cite{iCIPT2} (iterative CI\cite{iCI} with selection plus second-order perturbation\cite{SDS,SDSPT2}) should be even more promising, because
iCIPT2 is spin-symmetry adapted
and has the capability of targeting directly high-lying excited states that have little or even no overlap with the low-lying ones\cite{iVI,iVI-TDDFT}.


\section{Summary}\label{Conclusion}
\begin{figure}
\begin{tabular}{lcc}
(a)&\includegraphics[width=0.6\textwidth]{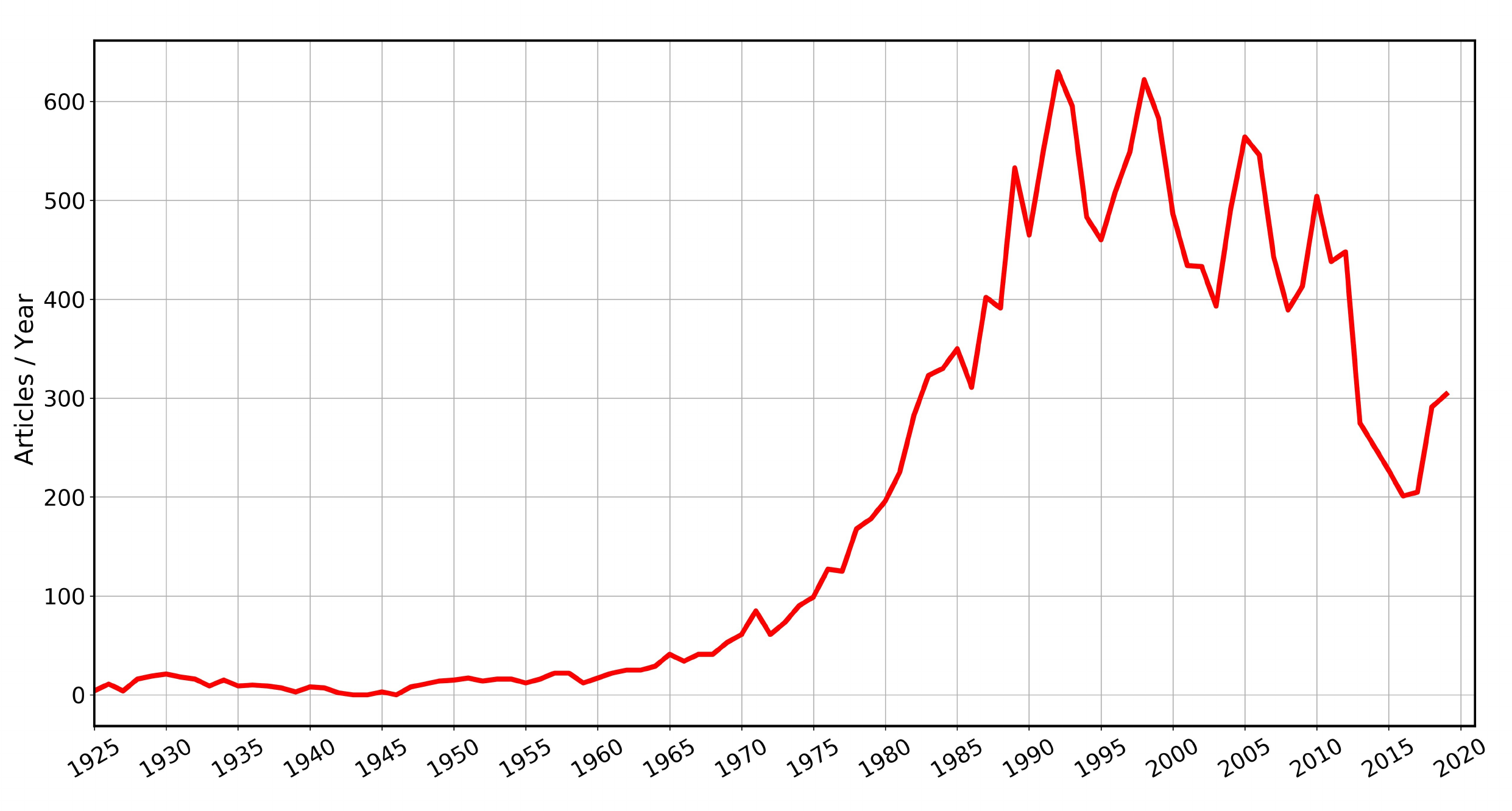}\\
(b)&\includegraphics[width=0.6\textwidth]{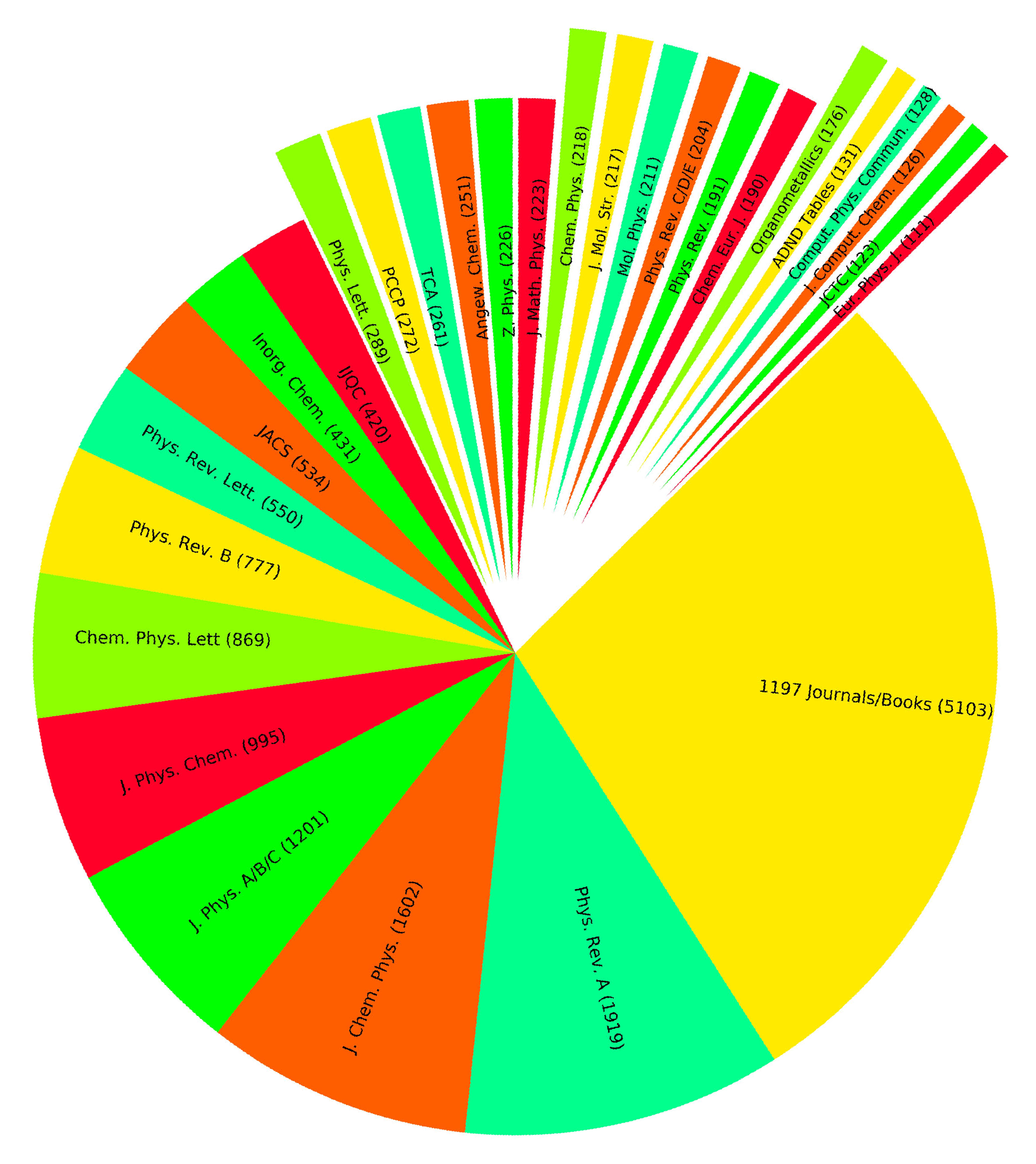}\\
\end{tabular}
\caption{(a) Number of relativistic articles per year; (b) Distributions of relativistic articles in journals.
JACS: J. Am. Chem. Soc.; IJQC: Int. J. Quantum Chem.; PCCP: Phys. Chem. Chem. Phys.; TCA: Theor. Chem. Acc.;
JCTC: J. Chem. Theory Comput.
}
\label{RTAMfig}
\end{figure}

Ironically in history, just one year after he proposed the famous relativistic equation of motion\cite{Dirac1928a,Dirac1928b}, Dirac himself stated\cite{DiracStatement}
that `relativistic effects are of no importance in the consideration of atomic and molecular structure and ordinary chemical reactions'.
Unfortunately, such a naive point of view lasted for nearly half a century, until the mid-1970s when relativistic effects\cite{pyykko1978relativistic} were found to be indeed very important for electronic structure, sometimes even of light atoms. Since then the field of relativistic quantum chemistry
has witnessed fast development, especially in the last 15 years,
as evidenced by nearly 20,000 relativistic articles\cite{PyykkoRTAM} (see Fig. \ref{RTAMfig}) as well as
more than 10 relativistic books\cite{Dyallbook,schwerdtfeger2002relativistic,schwerdtfeger2004relativistic,hess2003relativistic,hirao2004recent,Grantbook,barysz2010relativistic,kaldor2013theoretical,Reiherbook,dolg2015computational,Liubook}. With the advent of powerful
computational software\cite{ADFrev,BDF1,BDF2,BDF3,BDFrev2020,BERTHA,UTChem,NTChem,COLOGNE15,TURBOMOLE,DIRAC19,ReSpect,RAQET,BAGEL,Chronus-WIREs},
it can be envisaged that relativistic quantum chemistry will play an increasingly important role in the exploration of molecular science.
Apart from further improvement in the computational efficiency,
the most important and urgent methodological developments include (1) combination of the QED$@$Q4C/X2C and sf-X2C+so-DKH1 Hamiltonians with
sophisticated, symmetry-adapted wave function-based no-pair correlation methods (e.g., iCIPT2) for high-precision calculations of electronic structure
and (2) full implementation of the eQED Hamiltonian to establish the field of ``molecular QED'' for ultrahigh-precision calculations of
spectroscopic parameters. Here, efficient implementation of the frequency-dependent Breit integrals, which scale formally as the 8th power of the basis-set size, must first be accomplished. Works along these directions are being carried out in our laboratory.

\section*{Acknowledgement}
This research was financially supported by National Natural Science Foundation of China (Grant Nos. 21833001 and 21973054).

\section*{Data Availability Statement}
The data that supports the findings of this study are available within the article.

\clearpage
\newpage

\bibliographystyle{apsrev4-1}
\bibliography{BDFlib}

\end{document}